\documentclass[fleqn,usenatbib]{mnras}

\usepackage{newtxtext,newtxmath}

\usepackage[T1]{fontenc}
\usepackage{ae,aecompl}

\usepackage{graphicx}	
\usepackage{amsmath}	
\usepackage{xcolor}
\usepackage{ulem}
\usepackage{cancel}

\graphicspath{ {./Figs/} }

\newcommand{\cs}{c_{\rm s}}

\title[Rapid Formation of Super-Earths Around Low-Mass Stars]{Rapid Formation of Super-Earths Around Low-Mass Stars}

\author[Zawadzki et al.]{
Brianna Zawadzki,$^{1,2}$ \thanks{E-mail: bjz16@psu.edu}
Daniel Carrera,$^{3,1,2}$
Eric Ford$^{1,2,4}$
\\
$^{1}$Department of Astronomy and Astrophysics, 525 Davey Lab, The Pennsylvania State University, University Park, PA 16802, USA\\
$^{2}$Center for Exoplanets \& Habitable Worlds, 525 Davey Lab, The Pennsylvania State University, University Park, PA 16802, USA\\
$^{3}$Department of Physics and Astronomy, Iowa State University, Ames, IA, 50010, USA\\
$^{4}$Institute for Computational \& Data Sciences, The Pennsylvania State University, University Park, PA 16802, USA\\
}

\date{Accepted XXX. Received YYY; in original form ZZZ}

\pubyear{2021}

\begin{document}
\label{firstpage}
\pagerange{\pageref{firstpage}--\pageref{lastpage}}
\maketitle


\begin{abstract}
NASA's \textit{TESS} mission is expected to discover hundreds of M dwarf planets. However, few studies focus on how planets form around low-mass stars. We aim to better characterize the formation process of M dwarf planets to fill this gap and aid in the interpretation of \textit{TESS} results. We use ten sets of N-body planet formation simulations which vary in whether a gas disc is present, initial range of embryo semi-major axes, and initial solid surface density profile. Each simulation begins with 147 equal-mass embryos around a 0.2 solar mass star and runs for 100 Myr. We find that planets form rapidly, with most collisions occurring within the first 1 Myr. The presence of a gas disc reduces the final number of planets relative to a gas-free environment and causes planets to migrate inward. We find that roughly a quarter of planetary systems experience their final giant impact inside the gas disc, suggesting that some super-Earths may be able to reaccrete an extended gaseous envelope after their final giant impact, though these may be affected by additional process such as photoevaporation. In addition, we find that the final distribution of planets does not retain a memory of the slope of the initial surface density profile, regardless of whether or not a gas disc is present. Thus, our results suggest that present-day observations are unlikely to provide sufficient information to accurately reverse-engineer the initial distribution of solids.

\end{abstract}





\section{Introduction}

NASA's \textit{TESS} mission is expected to discover thousands of transiting exoplanets in the solar neighborhood. Simulated planet yields from \citet{Huang_2018} predict that \textit{TESS} will find about $10^{4}$ planets, 3500 of which will be Neptune size or smaller. Earlier simulations from \citet{Sullivan_2015} anticipate similar yields and predict that most of the planets detected from 2-minute cadence data of target stars will be super-Earths or sub-Neptunes. Since \textit{TESS} observes each sector of the sky for 27 days, it primarily targets short-period exoplanets. In addition, \textit{TESS} has a wide, red-optical bandpass which makes it more sensitive to planets transiting cool, low-mass stars such as M dwarfs than older missions like \textit{Kepler}. Models from \citet{Barclay_2018} suggest that \textit{TESS} is likely to find about 500 planets orbiting M dwarfs, more than the \textit{Kepler} mission by approximately an order of magnitude \citep{Huang_2018}. Furthermore, \textit{TESS} targets more nearby stars than \textit{Kepler}, opening up the door for radial velocity follow-up of \textit{TESS} planets. Previous studies that investigate the final assembly of planetary systems have understandably focused on systems around Sun-like stars; such systems have constituted most of the available data. However, the \textit{TESS} mission provides a compelling opportunity to explore planet formation around M dwarfs.

Protoplanetary discs consist of gas and dust that provide the initial conditions for planet formation. Disc properties vary with the spectral type of the central star, implying that planet formation is probably not uniform across all spectral types \citep{Andrews_2013}. This motivates studies of planet formation around M dwarfs.

Regardless of stellar spectral type, the first stage of planet formation is the the coagulation of dust grains into larger particles. As dust grains grow larger, the collision speeds $\Delta u_{\rm coll}$ increase \citep{Weidenschilling_1984}
\begin{equation}
    \Delta u_{\rm{coll}} \sim \cs \sqrt{\alpha \tau},
\end{equation}

\noindent where $\cs$ is the isothermal sound speed, $\alpha$ is the Shakura-Sunyaev turbulence parameter \citep{Shakura_1973}, and $\tau$ is the dimensionless friction time. Close to the star, particle growth is likely to be fragmentation-limited \citep{Birnstiel_2012}, so that the particle size is set by the fragmentation speed $u_{\rm{frag}}$ of the grain material.

Because these parameters vary by location in the disc, the maximum grain size is dependent on where in the disc grains form. The inner disc is largely dominated by fragmentation, leading to a dust surface density that goes as $r^{-1.5}$, while the outer disc is dominated by radial drift which cannot sustain high enough relative velocities for fragmentation over the disc lifetime \citep{Birnstiel_2012}. Our simulations focus on planet formation in the inner disc of an M dwarf, so this study includes models in which the embryos are initially distributed according to an $r^{-1.5}$ surface density profile. However, the distribution of solids may be rearranged as dust grains are converted into planetesimals. In addition to the fragmentation-limited case, we also consider the case where the solid distribution of embryos mirrors that of the turbulent gas disc and are initially distributed according to an $r^{-0.6}$ surface density profile. We investigate whether the resulting planetary systems can be used to distinguish between these two initial surface density profiles. One distinguishing feature of this study is that our simulations take the early stellar evolution of the M dwarf into account, including calculations for the changing stellar mass accretion rate and disc parameters. This is particularly important for M dwarfs due to the rate and length of their contraction onto the main sequence, but is not commonly accounted for in studies of M dwarf planet formation.

Aerodynamic drag causes the dust to drift radially inward in the turbulent disc \citep{Weidenschilling_1977}. As the solids continue to orbit, this radial drift combined with the turbulent motions in the disc cause particles to collide and either stick or fragment, depending on the collision speed and grain size \citep{Guttler_2010, Zsom_2010}. Particle growth stalls at around cm-size pebbles, meaning some other process is likely needed for continued growth \citep{Blum_2018}. Currently, the most promising mechanism to concentrate particles is a radial convergence of particle drift known as the streaming instability. The streaming instability causes particles to concentrate into long, dense azimuthal filaments which can gravitationally collapse into planetesimals \citep{Youdin_2005, Johansen_2007,Bai_2010a, Carrera_2015,Nesvorny_2019,Carrera_2020}. Another aerodynamic process that may play an important role is the concentration of particles inside turbulent eddies \citep{Cuzzi_2001, Cuzzi_2008, Estrada_2016, Hartlep_2017}.

As planetesimals collide to form embryos, they experience runaway growth followed by oligarchic growth. During runaway growth, the mass growth of planetesimals is given by $\dot M / M \propto M^{1/3}$ \citep{Greenberg_1978, Wetherill_1989}. When a small number of embryos become large enough to dominate the neighboring planetesimals, the system enters a stage of oligarchic growth. During this stage, growth is slower ($\dot M / M \propto M^{-1/3}$) and proceeds until the oligarchs reach their isolation mass \citep{Kokubo_1998, Kokubo_2000, Thommes_2003, Chambers_2006}.

The gas in the disc exerts torques on the embryos which can cause planets to migrate. There are two principal modes of planet migration:

\begin{description}
\item[\textbf{Type I:}]  Small planets embedded in the disc experience Type I migration. In it, the planet forms spiral density waves associated with its Lindblad resonances, and causes nearly co-orbital gas to follow horseshoe orbits. The associated overdensities lead to negative Lindblad torques and positive co-rotation torques, respectively \citep{Papaloizou_2000,Tanaka_2002,Paardekooper_2010,Paardekooper_2011,Coleman_2014}.

\item[\textbf{Type II:}]  When a planet is massive enough to open a gap in the disc, the co-rotation torque is all but eliminated and the Lindblad torque is reduced. The planet becomes coupled to the viscous evolution of the disc and orbital migration greatly slows down \citep{Lin_1986,DAngelo_2003,DAngelo_2008,Kley_2012}.
\end{description}

For our work we are primarily interested in Type I migration. Typically the Lindblad torque dominates, but depending on the background temperature gradient, the co-rotation torque may be able to overpower the Lindblad torque, in which case the planet experiences outward migration \citep{Paardekooper_2010}. There exists a regime in which Neptune-sized planets experience enhanced co-rotation torques, increasing the chance of outward migration. However, for smaller planets the co-rotation torque often saturates, causing the Lindblad torque to dominate and drive inward Type I migration \citep{Paardekooper_2011, Coleman_2014}. A more detailed description of disc torques is provided in Appendix \ref{app:torques}.

Type I migration stops because the gas disc does not extend all the way to the star; the star's magnetic field disrupts the disk out to the Alfv\'en radius \citep[e.g.][]{Long_2005}, where the magnetic energy density equals the kinetic energy density of the disk material. However, the location of the inner edge of the disc is not well-constrained, especially for low-mass stars. Keck interferometric observations of FU Ori suggest a disc inner edge of $0.07 \pm 0.02$ AU \citep{Eisner_2011}. Models based on the spectral energy distribution of TW Hya imply a disc inner edge of 0.02 AU \citep{Calvet_2002}, while Keck interferometric observations of TW Hya find a disc inner edge of $0.06 \pm 0.01$ AU \citep{Eisner_2006}. In this study, we choose a gas disc inner edge of 0.03 AU for most runs. However, we also compare our main results with simulations that used a gas disc inner edge of 0.01 AU in Section \ref{sec:results}.

This paper is organized as follows. Section \ref{sec:methods} introduces our simulation models. In Section \ref{sec:methods:disc} we describe the disc model and present relevant equations for the mass accretion rate, density profile, and gas torques. We detail the specific initial conditions of each of our models in Section \ref{sec:methods:inits}. Our main results are described in Section \ref{sec:results}, followed by a discussion of their significance and implications in Section \ref{sec:discuss}. Finally, we present our conclusions in Section \ref{sec:conclude}.

\section{Methods}
\label{sec:methods}

We run N-body simulations using the {\sc Mercury} code with the hybrid integrator \citep[]{Chambers_1999}. Some of our simulations are pure N-body, but most are N-body with the inclusion of an additional user-defined force that represents the effects of torques from a gas disc (i.e., inclination/eccentricity damping and migration) on each embryo.

All of our simulations take place around a 0.2 \(M_\odot\) star. We use the protoplanetary disc model of \citet{Ida_2016} and assume a quasi-steady state accretion disc with an evolving mass accretion rate. We assume the inner edge of the gas disc is located at 0.03 AU. Following \citet{Lenz_2019}, we assume planetesimals form early. For the purposes of setting the accretion rate, we assume that our N-body simulations begin when the disc is at the age of 1 Myr. We also assume that planetesimals have moved very little during the first 1 Myr (i.e., prior to the start of our simulation), from when they accumulated their ``initial'' mass.

\subsection{Disc Model}
\label{sec:methods:disc}

\begin{figure}
\centering
\includegraphics[width=\linewidth]{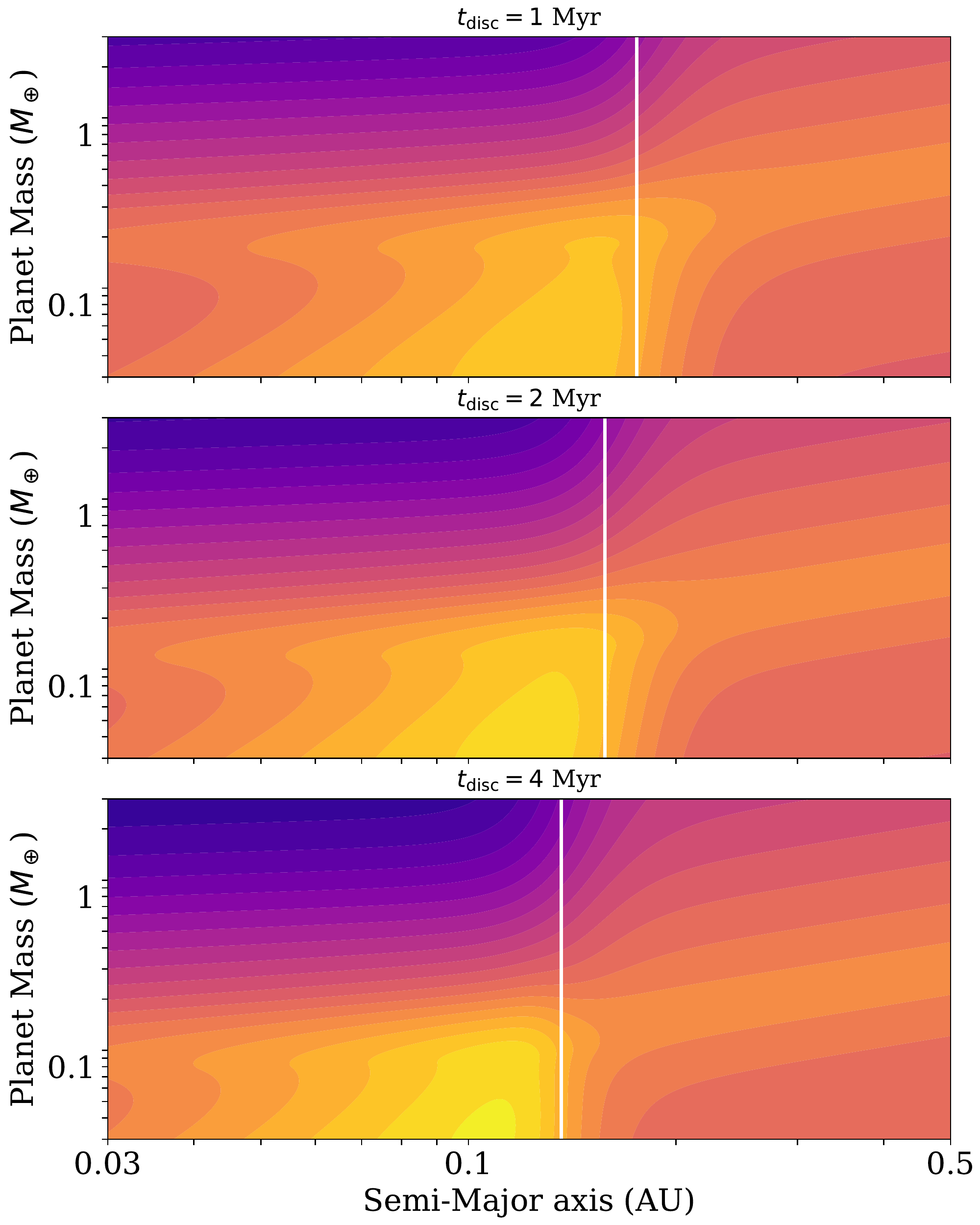}
    \begin{flushright}
    \includegraphics[width=7.65cm]{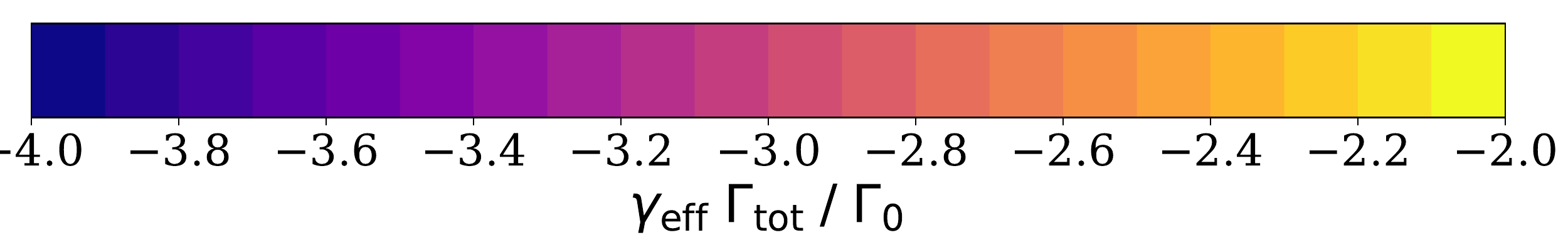}
    \end{flushright}
\caption{Lindblad torques and co-rotation torques act on embryos in the presence of a gas disc. Lindblad torques typically dominate, driving inward migration. Note that the entirety of the colorbar corresponds to negative values of $\gamma_{\rm eff} \Gamma_{\rm tot} / \Gamma_{0}$, meaning that there is a negative total torque on each embryo in the relevant ranges of planet mass and semi-major axis. Thus, planets in our simulations only experience inward migration. The vertical white line in each panel represents the distance where the dominant source of heating switches from viscous to irradiative \citep{Ida_2016}.}
\label{fig:torques}
\end{figure}

For a steady state accretion disc, the mass accretion rate is
\begin{equation}\label{eqn:accretion}
\dot{M} = 3 \pi \nu \Sigma,
\end{equation}

\noindent where $\dot{M}$ is the accretion rate, $\nu$ is the disc viscosity and $\Sigma$ is the gas surface density. We model $\nu$ as an $\alpha$-viscosity \citep{Shakura_1973},
\begin{equation}\label{eqn:viscosity}
    \nu = \alpha H c_{\rm s},
\end{equation}

\noindent where $H$ is the gas scale height, $c_{\rm s}$ is the speed of sound, and $\alpha$ is the Shakura-Sunyaev turbulence parameter. We follow \citet{Ida_2016} and set $\alpha = 10^{-3}$. The gas scale height is related to the sound speed by $H = c_{s} / \Omega$, where $\Omega$ is the Keplerian frequency. The isothermal sound speed is
\begin{equation}
    c_{s}= \sqrt{\frac{k_{B} T}{\mu m_{H}}},
\end{equation}
\noindent where $k_{B}$ is the Boltzmann constant, $\mu$ is the mean molecular weight of the gas, and $m_{H}$ is the mass of a hydrogen atom.

For our models, we use the accretion rate from \citet{Hartmann_2018},
\begin{equation}\label{eqn:hartmann}
\dot{M}=1.8 \times 10^{-8} M_{0.7}^{2} t_{6}^{-1} \; \mathrm{M}_{\odot} \mathrm{yr}^{-1},
\end{equation}
\noindent where $M_{0.7}$ is the mass as $M_{\star}/0.7 M_{\odot}$ and $t_{6}$ is the stellar age in Myr. For a $0.2 M_{\odot}$ star at $1$ Myr we get $\dot{M}_{\star} = 1.5\, \rm{x} \, 10^{-9}$ $M_{\odot}$/yr. 

Following \citet{Ida_2016}, we distinguish between a viscous regime where viscous heating dominates, and an irradiative regime where irradiation from the central star dominates. For the latter regime we need an expression for the star's luminosity. We calculate the luminosity with stellar evolution tracks from \citet{Marigo_2017}. For example, a 1 Myr old star with a mass of 0.2 \(M_\odot\) and a metallicity of $Z = 0.01$, would have a luminosity of $L_{0} = 0.26 \: L_\odot$.

Plugging in a stellar mass of $0.2 M_{\odot}$ and $\alpha = 10^{-3}$, the temperature profiles from \citet{Ida_2016} become
\begin{equation}
    T_{\mathrm{vis}} \simeq 123.4 \: \left( \frac{\dot{M}_{\star}}{10^{-8} M_{\odot} / \mathrm{yr}} \right)^{2 / 5}\left(\frac{r}{1\: \mathrm{au}}\right)^{-9 / 10} \mathrm{K}
\end{equation}

\noindent for the viscous heating regime and 
\begin{equation}
    T_{\mathrm{irr}} \simeq 188.8\: \left( \frac{L_{\star}}{1 L_{\odot}} \right)^{2 / 7}\left(\frac{r}{1\: \mathrm{au}}\right)^{-3 / 7} \mathrm{K}
\end{equation}

\noindent for the irradiative heating regime. For a 1 Myr old disc, the boundary between the viscous and irradiative regimes occurs at a distance of $r = 0.18$ AU.

We smooth the transition between the viscous and irradiative regimes so the temperature profile is differentiable at the boundary, allowing continuous torque calculations. We smooth the viscous-irradiative boundary with
\begin{equation}\label{eqn:smooth}
T(r) = \left( T_{\mathrm{vis}}(r)^{n} + T_{\mathrm{irr}}(r)^{n} \right)^{1/n},
\end{equation}
where we arbitrarily select $n=20$. Our simulations take place mostly in (inner) viscous regime, where the gas surface density is given by

\begin{equation}\label{eqn:ida_sig}
\Sigma_{\mathrm{g}, \mathrm{vis}} \simeq 1522 \: \left( \frac{\dot{M}_{\star}}{10^{-8} M_{\odot} / \mathrm{yr}} \right)^{3 / 5}\left(\frac{r}{1\: \mathrm{au}}\right)^{-3 / 5} \mathrm{g\: cm}^{-2}.
\end{equation}

\noindent The viscous-irradiative boundary moves inward as the simulation progresses due to its dependence on stellar luminosity and disc accretion rate, which also change with time.

\subsection{Disc Torques}
\label{sec:methods:torque}

In addition to the gravitational N-body interactions, our planetary embryos experience disk torques. The Lindblad torque arises from spiral density waves and is typically negative, causing planets to migrate inward. The co-rotation torque is generated by gas in horseshoe orbits around a planet. As the gas changes temperature the resulting density gradient creates a positive torque. The total torque felt on each embryo may be described as the sum of Lindblad and co-rotation torques.
\begin{equation}\label{eqn:tot_torque}
    \Gamma_{\mathrm{tot}}=\Gamma_{\mathrm{Lind}}+\Gamma_{\mathrm{CoRot}}
\end{equation}

\noindent The negative Lindblad torque usually dominates. As a result, $\Gamma_{\mathrm{tot}}$ is usually negative, and embryos experience inward migration. Figure~\ref{fig:torques} shows our model disc torques for a range of semi-major axes and planet masses. The relevant formulas and details regarding the disc torques are included in Appendix \ref{app:torques}.

\subsection{Initial Conditions}
\label{sec:methods:inits}

\begin{figure*}
\centering
\includegraphics[width=\linewidth]{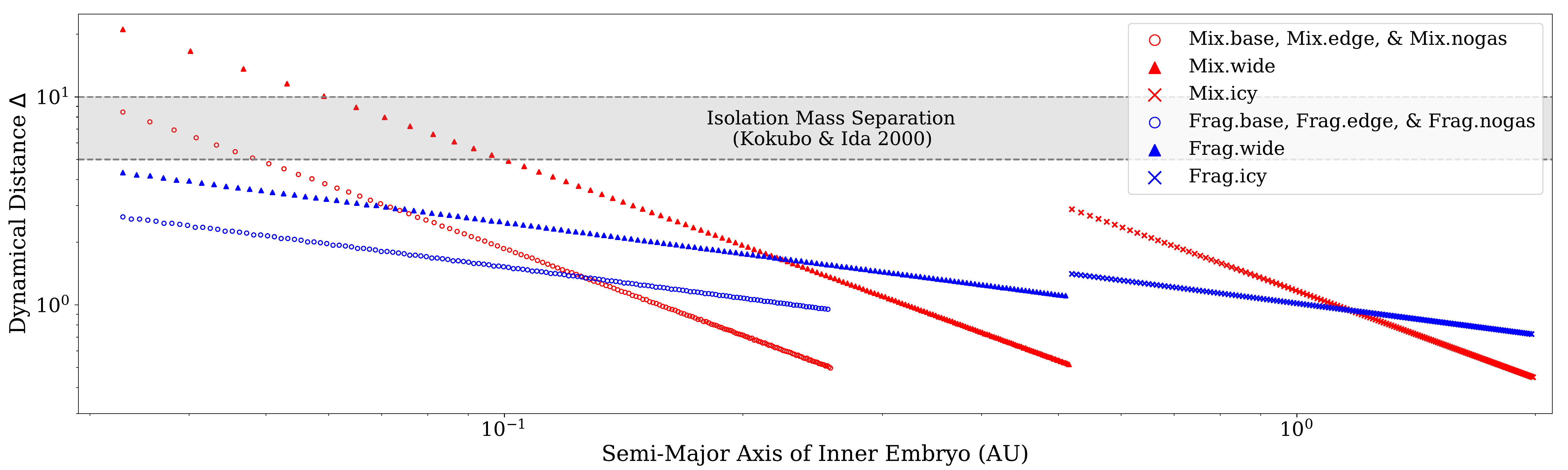}
\caption{The initial dynamical distance between embryos decreases with increasing semi-major axis for each of our models. Following \citet{Kokubo_2000}, we consider $5-10 R_{\mathrm{Hill}}$ to be the range of dynamical distances where the embryos are dynamically isolated from other planetesimals. Most of our models actually begin at lower values of $\Delta$, increasing the resolution of our simulations. In Mix.wide, the 5 innermost embryos have a initial $\Delta \gtrapprox 10$. Because this is only 5 out of 147 embryos, and Mix.wide yielded similar outcomes compared to the other models, we conclude that the presence of a few low-resolution embryos does not negate the validity of the model.\label{fig:dynamical}}
\end{figure*}

We consider ten models and run a set of 100 N-body simulations for each. Each model differs in solid surface density profile, initial embryo semi-major axis range, whether a gas disc is present during formation, and the location of the inner edge of the gas disc. Table \ref{tab:models} summarizes these differences.

We assume that planetesimals form early ($t = 0.1$ Myr) in all models so that the initial distribution of solids reflects that of the young disc \citep{Lenz_2019}. Early in the disc lifetime, the boundary between the viscous and irradiative regimes occurs near the outer edge of our initial embryo distribution for most models. Thus, the majority of the embryos are initialized at distances that are dominated by viscous heating. All of our simulations initially arrange the embryo semi-major axes so that the solid surface density follows a uniform power law,
\begin{equation} \label{eqn:powerlaw}
\Sigma = \Sigma_{0} \; r^{-\gamma}.
\end{equation}

\noindent A uniform power law is advantageous as it allows for better characterization of the change between the initial and final solid surface density distribution (Section \ref{sec:results:inits}). Since we assume that planetesimals form early, the solid mass needs to reflect the mass available when the disc was younger and more massive. Our N-body simulations begin at t = 1 Myr, well after the epoch of planetesimal formation ($\sim 0.1$ Myr). Since $\dot{M}$ is inversely proportional to $t$ (Equation \ref{eqn:hartmann}) and $\dot{M}$ is proportional to $\Sigma$ (Equation \ref{eqn:accretion}), the disc mass in solids at the epoch of planetesimal formation ($t \sim 0.1$ Myr) is 10 times greater than at the start of the N-body portion of our simulations (t = 1 Myr). In addition, dust evolution is likely to enhance the dust-to-gas ratio in the inner disc by up to a factor 10 \citep{Birnstiel_2012}. In the end, enhancement in the solid-to-gas ratio caused by gas accretion and dust migration is uncertain. Our initial conditions imply an enhancement of $9-25\times$, which should be in the correct magnitude range when the simulations begin at $t_0 = 1$ My. This model corresponds to a disc mass of $0.024 M_\odot$ out to 20 AU at t = 1 Myr.

All simulations begin with 147 embryos, each with an initial mass of $0.08 M_\oplus$. The initial inclination and eccentricity of each embryo are 0.1 degrees and 0.01, respectively. The initial argument of pericentre, longitude of the ascending node, and mean anomaly are all randomly generated from 0 to $2\pi$ for each embryo. 

Following \citet{Birnstiel_2012}, we distinguish between two formation scenarios:
\begin{enumerate}
    \item If dust particles are well-mixed with the gas (models labeled `Mix') the solid surface density profile will mirror the $r^{-0.6}$ power law of the viscously heated inner disc \citep{Ida_2016}.
    \item If dust sizes are fragmentation-limited (models labeled `Frag'), aerodynamic drag can rearrange the solids into a steeper $r^{-1.5}$ density profile \citep{Birnstiel_2012}.
\end{enumerate}

\noindent As a result, we have
\begin{equation}\label{eqn:mixpl}
\Sigma_{\mathrm{mix}} = \Sigma_{0, \mathrm{mix}} \; \left(\dfrac{r}{\mathrm{AU}} \right)^{-0.6}
\end{equation}

\noindent for Mix models, and
\begin{equation}\label{eqn:frag}
\Sigma_{\mathrm{frag}} = \Sigma_{0, \mathrm{frag}} \; \left(\dfrac{r}{\mathrm{AU}} \right)^{-1.5}
\end{equation}

\noindent for Frag models. The value of $\Sigma_{0}$ is dependent on the total mass of the system, the power law, and the locations of the inner and outer edges of the embryo belt. All ten of our models have the same total mass, so values of $\Sigma_{0}$ differ between models with different initial power laws or embryo semi-major axis ranges. Note that the presence or absence of a gas disc does not affect the value of $\Sigma_{0}$.

Our base models, Mix.base and Frag.base, are N-body simulations in which embryos are initialized from 0.033 AU to 0.26 AU. Embryos experience Type I migration as well as damping of eccentricity and inclination from interactions with the gas disc, which has its inner edge at 0.03 AU.

Each of the other models varies one parameter relative to the corresponding base model. Models Mix.wide and Frag.wide feature a wider initial embryo belt, extending the outer limit of the base models by a factor of two to initialize embryos from 0.033 AU to 0.52 AU.

Models Mix.icy and Frag.icy push the embryo belt even further out so that embryos are beyond the ice line at the beginning of the simulation, ranging from 0.52 AU to 2.0 AU. We assume the ice line is located where the disc temperature is T=170 K. At a disc age of 1 Myr, this corresponds to an ice line at 0.52 AU. The 2.0 AU outer bound of the icy model embryo belt was chosen to give the embryos comparable initial dynamical distances compared to the base models (see Figure \ref{fig:dynamical}). These runs are the most computationally expensive and due to limited computing power we run only 50 simulations each for Mix.icy and Frag.icy.

Current observations do not tightly constrain the location of the inner edge of a gas disc, especially around M dwarfs. While most models place have the inner edge of the gas disc located at 0.03 AU, models Mix.edge and Frag.edge move the edge inward to 0.01 AU. Embryos are placed at the same semi-major axes as in the base models.

Lastly, models Mix.nogas and Frag.nogas are pure N-body simulations. While the other models interact with the gas disc, the gas-free models do not include the torques caused by the presence of gas. Note that in every model, embryos grow due to collisions with other embryos and the total mass of the system is held constant.

\begin{table*}
  \centering
  \caption{The initial conditions for our ten sets of simulations. For each model, the initial number of embryos $N_{\rm emb} = 147$ and the initial mass of each embryo $m_{\rm emb} = 0.08 M_\oplus$, so the total mass in each simulation is held constant. The central star has mass $M_\star = 0.2 M_\odot$. Mix models correspond to a formation scenario where the dust particles that gave rise to planetesimals are assumed to have been well-mixed with the gas, while Frag models correspond to a fragmentation-dominated distribution. The value of $\gamma$, the power-law index of the solid density profile, is dependent on whether the dust is well coupled to the gas ($\gamma = 0.6$), or whether the distribution is fragmentation-dominated ($\gamma = 1.5$). The presence of a gas disc indicates that Type I migration torques act on the embryos. The location of the inner edge of the gas disc and the initial location of embryos vary by model. Nearly all the embryos in each model are initially closer than $5-10 R_{\mathrm{Hill}}$, the typical separation at the end of oligarchic growth where embryos become dynamically isolated from each other \citep{Kokubo_2000}. Because embryos begin dynamically close together, more gravitational interactions occur which increases the effective resolution of the simulation. However, the 5 innermost Mix.wide embryos are spaced more widely than $10 R_{\mathrm{Hill}}$ due to the semi-major axis range and $r^{-0.6}$ solid density profile. These embryos are the exception, and because nearly all embryos begin below $10 R_{\mathrm{Hill}}$, we conclude that this does not impact the validity of the model. The median $\Delta$ for each model is of order unity for all models.}
  \label{tab:models}
  \begin{tabular}{ccccccccc}
  Model Name & $\gamma$ & Gas? & Disc Edge/AU & $a_{\rm init}$/AU & $\Sigma_{0}$/$\mathrm{g}\;\mathrm{cm}^{-2}$ & Innermost $\Delta$ & Outermost $\Delta$ & Median $\Delta$\\ 
  \hline

  Mix.base  & 0.6 & Yes & 0.03 & $0.033 - 0.26$ & 4.88 & 8.449 & 0.497 & 0.946 \\
  Mix.wide  & 0.6 & Yes & 0.03 & $0.033 - 0.52$ & 1.78 & 21.114 & 0.518 & 1.011 \\
  Mix.icy   & 0.6 & Yes & 0.03 & $0.52  - 2.0 $ & 0.312 & 2.882 & 0.448 & 0.776 \\
  Mix.edge  & 0.6 & Yes & 0.01 & $0.033 - 0.26$ & 4.88 & 8.449 & 0.497 & 0.946 \\
  Mix.nogas & 0.6 & No  & N/A & $0.033 - 0.26$ & 4.88 & 8.449 & 0.497 & 0.946 \\
  \hline
  Frag.base  & 1.5 & Yes & 0.03 & $0.033 - 0.26$ & 0.760 & 2.648 & 0.952 & 1.401 \\
  Frag.wide  & 1.5 & Yes & 0.03 & $0.033 - 0.52$ & 0.462 & 4.324 & 1.108 & 1.761 \\
  Frag.icy   & 1.5 & Yes & 0.03 & $0.52  - 2.0 $ & 0.360 & 1.408 & 0.723 & 0.956 \\
  Frag.edge  & 1.5 & Yes & 0.01 & $0.033 - 0.26$ & 0.760 & 2.648 & 0.952 & 1.401 \\
  Frag.nogas & 1.5 & No  & N/A & $0.033 - 0.26$ & 0.760 & 2.648 & 0.952 & 1.401 \\

  \end{tabular}
\end{table*}
Let $\Delta$ be the separation between adjacent embryos in terms of their mutual Hill radius,

\begin{equation}\label{eqn:dyn_dist}
\Delta = \left(\frac{a_{2} - a_{1}}{R_{\mathrm{Hill}}}\right)
\end{equation}

\begin{equation}\label{eqn:hill_radii}
R_{\mathrm{Hill}} =  \left(\frac{m_{1} + m_{2}}{3 M_{\star}}\right)^{1 / 3}  \left(\frac{a_{1} + a_{2}}{2}\right)
\end{equation}\

\noindent where $m_{1}$ and $m_{2}$ are the embryo masses and $a_{1}$ and $a_{2}$ are their semi-major axes. Given equal-mass embryos, the separation between embryos must change in order to satisfy the model's solid surface density profile (Equations \ref{eqn:mixpl} and \ref{eqn:frag}). For Mix models, the initial dynamical distance between embryos ranges from $\Delta = 21.114$ at the inner edge of the embryo belt to $\Delta = 0.448$ at the outer edge. For Frag models. the initial dynamical distance ranges from $\Delta = 4.324$ at the inner edge of the embryo belt to $\Delta = 0.723$ at the outer edge. In other words, embryos that are initialized closer to the central star are dynamically farther apart than those that are initialized farther from the central star, which are closer together when measured in $R_{\mathrm{Hill}}$.

This information is summarized in both Figure~\ref{fig:dynamical} and Table~\ref{tab:models}. The isolation mass separation \citep{Kokubo_2000} is also shown in Figure~\ref{fig:dynamical}. Only a few Mix.wide embryos have an initial dynamical distance larger than $\Delta = 10$. In other words, our simulations begin before embryos have completed their growth into isolation-mass bodies and we resolve the final stages of embryo formation. Table~\ref{tab:models} also provides the median initial value of $\Delta$ for each model, which tends to be near $1 R_{\mathrm{Hill}}$ for all models.


\section{Results}
\label{sec:results}

\subsection{Example Run}
\label{sec:results:ex}

\begin{figure}
\centering
\includegraphics[width=\linewidth]{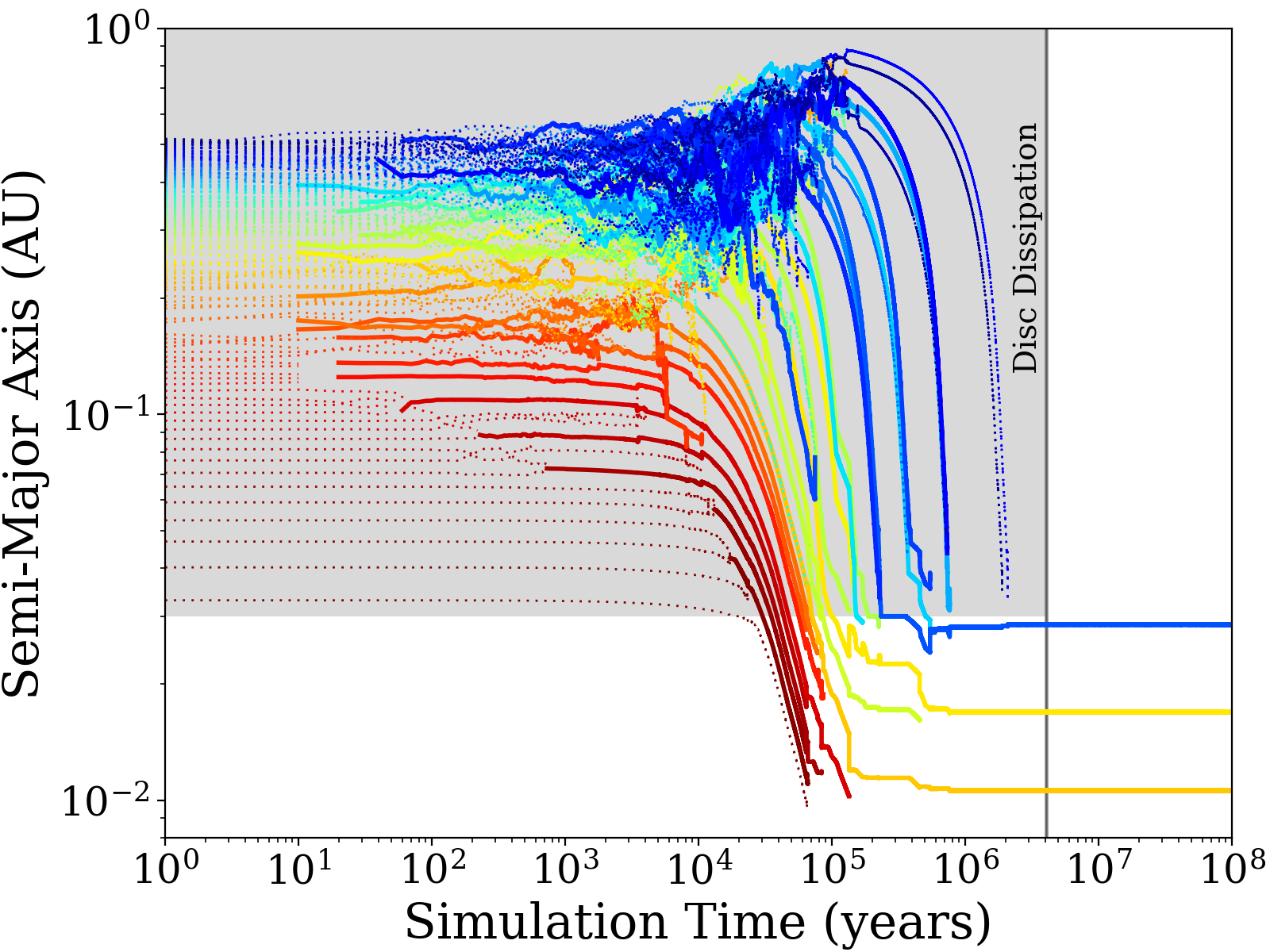}
\caption{Embryo semi-major axis over time for a typical simulation of model Mix.wide. Each line represents an embryo, and the abrupt end of a line means that embryo collided with another embryo. Line color corresponds to the initial semi-major axis of the embryo; the red lines correspond to the innermost embryos (near 0.033 AU), and the blue lines correspond to the outermost embryos (near 0.52 AU). Dotted lines represent embryos that have not yet collided, while thicker solid lines indicate that the embryo has collided at least once. The grey shaded region represents the gas disc. The inner edge of the gas disc lies at 0.03 AU until the disc experiences rapid exponential dissipation due to photoevaporation beginning at 5 Myr (4 Myr simulation time). This particular simulation results in several embryos locked in chains of mean motion resonance, which cause them to migrate into the disc cavity via inward Type I migration. After further gravitational interactions and collisions, 3 stable planets emerge after only a few Myr.}
\label{fig:semivstime}
\end{figure}

Figure \ref{fig:semivstime} shows the semi-major axis of planetary embryos as a function of simulation time for a typical Mix.wide simulation. Notice the rapid inward migration of the embryos between approximately $10^{4}$ and $10^{6}$ years. In this particular example, 3 final planets emerge. The planets settle in their final orbits several Myr before the disc undergoes rapid exponential dissipation due to photoevaporation. The outermost planet stops migrating at the inner edge of the gas disc.

In addition to showing the general presence of Type I migration in the simulations, Figure \ref{fig:semivstime} shows that planets migrate further inward than the innermost edge of the disc. This happens when planets become locked into chains of mean motion resonances. Although planets that are inside the disc cavity do not feel the disc torques directly, they can be in resonance with planets inside the gas disc. In this case, the resonance lock pushes inner planets further inward when an outer planet in the resonance chain experiences inward Type I torques \citep[e.g][]{Terquem_2007, Brasser_2018, Carrera_2018}. Thus, we see significant inward migration past the inner edge of the gas disc.

Figure \ref{fig:semivstime} also shows that collisions occur both inside and outside of the gas disc. In this specific run, 11.8\% of collisions occur outside of the gas disc (i.e., interior to 0.03 AU). We discuss the location of the collisions in Section \ref{sec:results:form}. Planets that form inside the gas disc may retain a gaseous envelope, even after inward migration. However, it is likely that most gaseous envelopes will later be removed by a late collision inside the disc cavity.

The three planets formed in this run have final masses of $2.5 \: \mathrm{M_{\oplus}}$, $5.1 \: \mathrm{M_{\oplus}}$, and $4.1 \: \mathrm{M_{\oplus}}$, ordered by increasing semi-major axis. The inner pair has a final dynamical separation of $\Delta = 12.78$, a mutual inclination of $1.03^{\circ}$, and a period ratio of $2.01$. The outer pair has a final dynamical separation of $\Delta = 15.24$, a mutual inclination of $0.50^{\circ}$, and a period ratio of $2.19$. Similar period ratios are common among \textit{Kepler} systems, which we further examine in Figure \ref{fig:kepler}. Each of the three planets have relatively low final eccentricities (innermost to outermost: 0.099, 0.022, and 0.014) and final inclinations (innermost to outermost: $0.60^{\circ}$, $0.43^{\circ}$, and $0.08^{\circ}$).

The angular momentum deficit (AMD) of a planetary system can be used as another measure of stability \citep{Laskar_2017, Petit_2017}. This system meets the criteria for AMD stability both at the end of the simulation (100 Myr) and as early as around 2 Myr simulation time, so this system quickly settled into an orbital arrangement that is likely to persist indefinitely.

These results are representative of a typical Mix.wide simulation. In Section \ref{sec:results:form} we generalize the results regarding the formation timescale of planets in all models. Section \ref{sec:results:planets} details the trends in planet characteristics for each model. Lastly, we discuss the insensitivity of the final planets to the model initial conditions in Section \ref{sec:results:inits}.

\subsection{Formation Timescale}
\label{sec:results:form}

\begin{figure*}
\centering
\includegraphics[width=\linewidth]{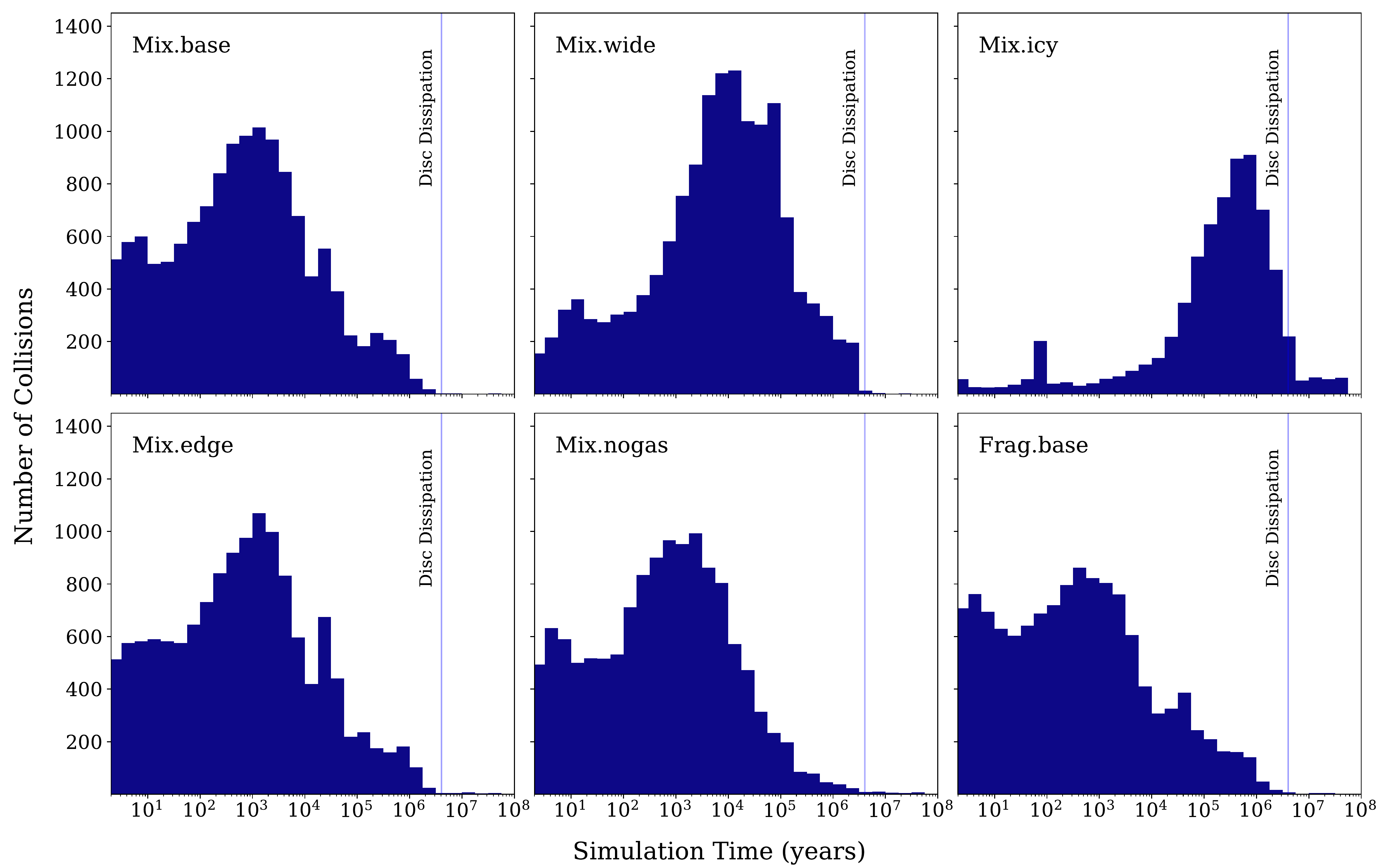}
\caption{The number of collisions for each Mix model with respect to time. The number of collisions for Frag.base is also provided for comparison, but Frag results do not differ significantly from Mix results. In general, planets form quickly, with embryo collisions peaking at approximately $10^{3}$ yr for models Mix.base, Mix.edge, and Mix.nogas, as well as the corresponding Frag models. Models Mix.wide and Mix.icy initialize the embryos across a wider range of semi-major axes, so the embryos require more time to collide. Collisions peak around $10^{4}$ yr for Mix.wide runs and between $10^{5}$-$10^{6}$ yr for Mix.icy runs, suggesting that even more widely dispersed embryos quickly form planets. Note that in most models, an insignificant number of collisions occur after the gas disc dissipates, indicating that late M dwarf planets largely form while the gas disc is still present. However, some collisions occur inside the disc cavity.}
\label{fig:collisions}
\end{figure*}

Embryos collide quickly, with nearly all collisions occurring within the disc lifetime. For models that initialize embryos out to 0.26 AU (base, edge, and nogas models), we find that collisions peak around $10^{3}$ years. Models that initialize embryos out to 0.52 AU (wide models) have a collision rate that peaks around $10^{4}$ years. This is likely due to the fact that embryos are initialized across a wider area; larger initial dynamical separations require more time for the collision rate to peak \citep{Chambers_1996,Smith_2009}. This effect is shown even more clearly in the icy models, which initialize embryos out to 2 AU. Figure \ref{fig:collisions} shows the collisions with respect to time for all models.

For all models, most collisions occur within the simulation's first 1 Myr. This means that planets form almost entirely before the gas disc dissipates, with few additional collisions occurring after dissipation. Averaged over all simulations, the first 1 Myr contained 96.1\% of collisions. For models with gas present, 98.7\% of collisions occur before the gas disc begins to rapidly dissipate 4 Myr into the simulation. Though some late collisions occur, planets largely form during the lifetime of the disc. Thus, the semi-major axis of the final impact may be an important factor when considering whether a planet has retained a gaseous envelope.

For models that include a gas disc, an average of 7.94\% of collisions occur inside the disc cavity. Models which initialized embryos over a smaller range (Mix.base, Frag.base, Mix.edge, and Frag.edge) had an average of 6.29\% of collisions occur inside the disc cavity, while the more extended models (Mix.wide, Frag.wide, Mix.icy, and Frag.icy) had an average of 10.15\% of collisions occur inside the disc cavity. The specific location of the inner edge of the gas disc did not have a large impact on the number of collisions inside the disc cavity; Mix.base had 7.14\% of collisions inside the disc cavity, while Mix.edge had 6.27\%. Similarly, Frag.base had 6.58\% of collisions inside the disc cavity, while Frag.edge had 5.16\%. Because gas torques cause embryos to migrate inward as the simulation progresses, late impacts are more likely to occur inside the disc cavity than early impacts. A planet may acquire a gaseous envelope from forming within the gas disc, but this atmosphere may later be entirely or partially removed by a late collision.

\begin{figure}
\centering
\includegraphics[width=\linewidth]{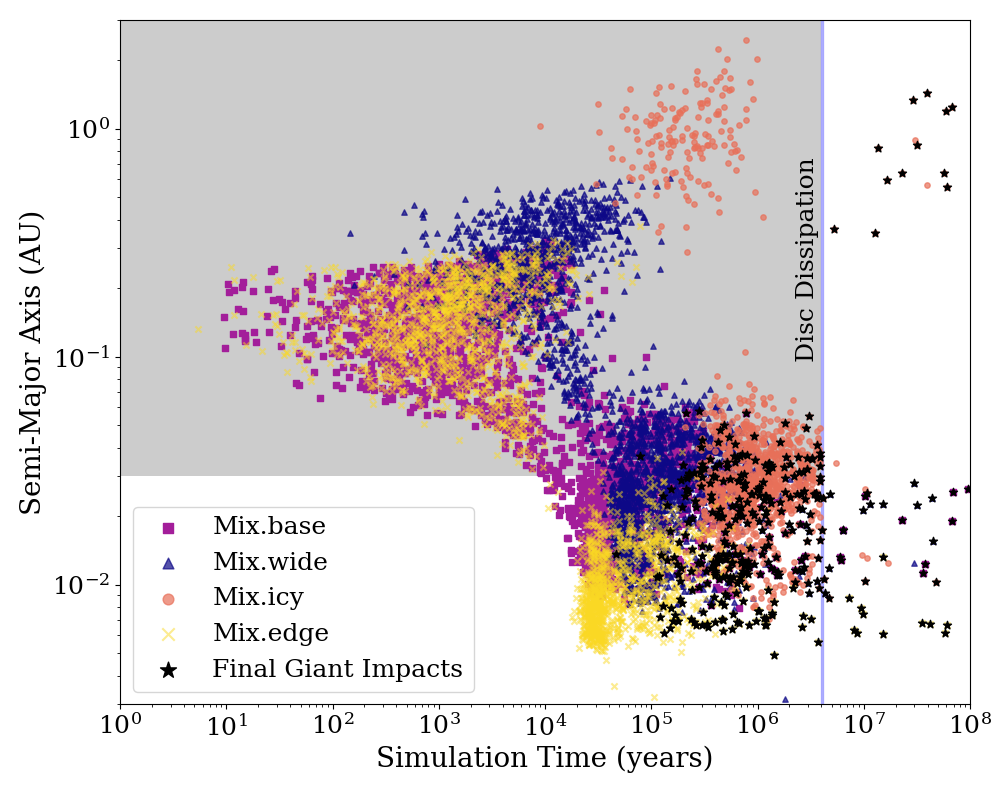}
\caption{The time and semi-major axis of all giant impacts (collisions where both $m_{1}+m_{2} > 0.5 M_{\oplus}$ and $m_{1}/m_{2} \leq 5$) in gaseous Mix models. There are two distinct clusters for each model: one for early giant impacts that occur before significant migration has occurred, and one for later impacts where the embryos have migrated towards or inside the disc cavity. The vertical blue line shows the time of disc dissipation, and the shaded gray area represents the times and locations where the gas disc exists. We also denote the time and location of the final giant impact of each run with a black star. Most gaseous Mix runs ($85.4\%$) have a final giant impact that occurs before disc dissipation. No final giant impacts from Mix.edge occur within the gas disc, as planets rapidly migrate into the disc cavity. Considering only Mix.base, Mix.wide, and Mix.icy, nearly a quarter ($24.4\%)$ of final giant impacts occur in the gas disc.}
\label{fig:giantimpacts}
\end{figure}

Let us define a \textit{giant impact} as a collision between two embryos where the total mass $m_{1}+m_{2} > 0.5 M_{\oplus}$ and the mass ratio $m_{1}/m_{2} \leq 5$, where $m_{1}$ is the more massive embryo. In models with gas, 16.9\% of all collisions are giant impacts. The time and location of giant impacts for gaseous Mix models is shown in Figure \ref{fig:giantimpacts}. The location of these impacts are largely dependent on how late the collision occurred; early giant impacts occur mostly where embryos are initially distributed, while later giant impacts occur further inward due to migration. Many late giant impacts occur inside the disc cavity. For all models, nearly all giant impacts occur while the gas disc is still present, but a few extremely late impacts did occur. Models without gas experienced fewer giant impacts in general, but the giant impacts that did occur were further out due to embryos not experiencing disc migration.

Figure \ref{fig:giantimpacts} also shows the final giant impact of each system for all Mix models including a gas disc. Of these final giant impacts, $17.4\%$ occurred in the gas disc before disc dissipation, $68.0\%$ occurred inside the disc cavity before disc dissipation, and the rest occurred after disc dissipation. However, none of the final giant impacts in the gas disc are contributed from Mix.edge runs; planets simply migrate into the disc cavity too rapidly. Excluding Mix.edge and considering only Mix.base, Mix.wide, and Mix.icy, nearly a quarter ($24.4\%$) of final giant impacts occur in the gas disc. The results are comparable for the corresponding Frag models: $21.1\%$ occurred in the gas disc before disc dissipation and $68.0\%$ occurred inside the disc cavity before disc dissipation. Excluding Frag.edge, $29.6\%$ of final giant impacts occur in the gas disc. Averaged over all gaseous models, only $12.8\%$ of final giant impacts occurred after disc dissipation.

\subsection{Planet Characteristics}
\label{sec:results:planets}

\begin{figure*}
\centering
\includegraphics[width=\linewidth]{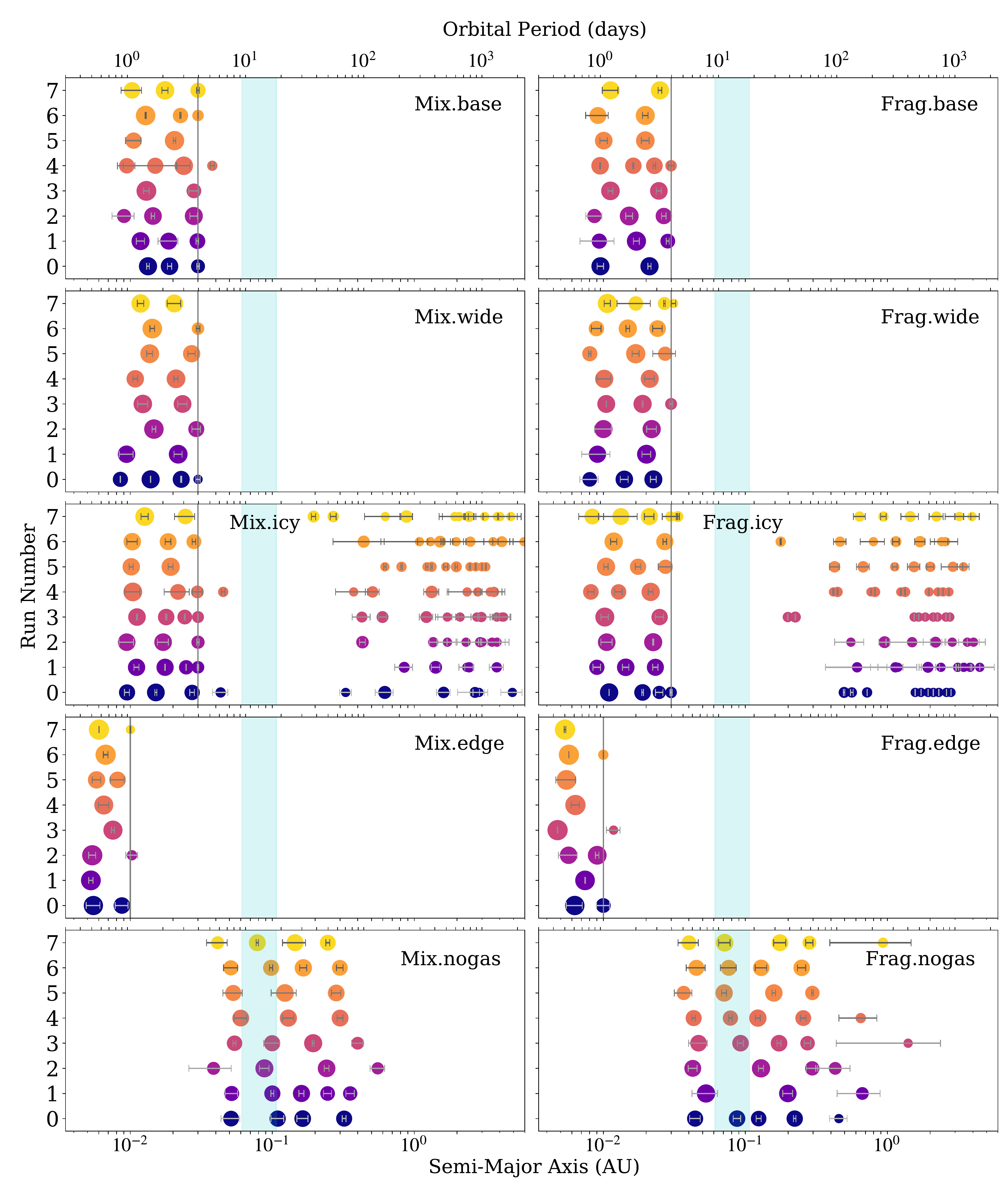}
\caption{For each model we display the final planets that result from the first 8 simulations. Each run is given its own row and color. The lower x-axes show the semi-major axis, while the top x-axes give the orbital period. The radius of each planet is approximated by $r_{p} \propto m_{p}^{1/3}$, and the horizontal lines denote the location of each planet's apastron and periastron. The vertical grey line in each panel represents the inner edge of the gas disc, if present. The vertical cyan band represents the habitable zone around the star, provided only for reference \citep{Kopparapu_2013}. Due to the torques shown in Figure \ref{fig:torques}, the inclusion of a gas disc creates planets that migrate inward toward the central star. A quantitative summary of the final planets in each model is provided in Table \ref{tab:planetsummary}.}
\label{fig:planets}
\end{figure*}

\begin{table*}
  \centering
  \caption{A summary of the final planet characteristics for each of our 10 models. Each value is computed from all simulations in the specified model. We ran 100 simulations for each model except the icy models, for which we ran 50 simulations each.}
  \label{tab:planetsummary}
  \begin{tabular}{cccccc}
  Model & Mean No.   & Median  & Maximum & Median Mutual & AMD-Stable \\
  Name  & Planets & Mass ($M_{\oplus}$) & Mass ($M_{\oplus}$) & Inclination & Systems at 100 Myr   \\
  \hline

  Mix.base  & 2.71  & 4.14  & 11.62 & $0.55^{\circ}$ & 96\%  \\
  Mix.wide  & 2.85  & 3.90  & 11.46 & $1.41^{\circ}$ & 93\%  \\
  Mix.icy   & 12.44 & 0.16  & 9.55  & $8.66^{\circ}$ & 6\%   \\
  Mix.edge  & 1.48  & 10.06 & 11.76 & $1.47^{\circ}$ & 100\% \\
  Mix.nogas & 4.19  & 2.63  & 7.80  & $7.44^{\circ}$ & 67\%  \\
  \hline
  Frag.base  & 2.88  & 4.18 & 9.47  & $1.01^{\circ}$ & 91\% \\
  Frag.wide  & 2.84  & 4.06 & 10.82 & $1.50^{\circ}$ & 94\% \\
  Frag.icy   & 11.84 & 0.16 & 9.71  & $0.54^{\circ}$ & 10\% \\
  Frag.edge  & 1.54  & 8.15 & 11.76 & $0.33^{\circ}$ & 98\% \\
  Frag.nogas & 4.30  & 2.71 & 7.64  & $5.96^{\circ}$ & 72\% \\

  \end{tabular}
\end{table*}

We find that both models that do not include a gas disc (Mix.nogas, Frag.nogas) produce planets with similar orbital characteristics (Figure \ref{fig:planets}). The final semi-major axes are comparable to the initial embryo distribution, except for some planets which were flung outward by close encounters. These simulations produced an average of 4.25 planets per simulation.

Planets resulting from models that include a gas disc (base, wide, icy, and edge models) look quite different than the gas-free models. The presence of a gas disc causes planets to migrate inward such that the outermost planet has a semi-major axis at or near the inner edge of the disc (at 0.03 AU or 0.01 AU, depending on the model). Most planets do not experience any gravitational perturbations after the disc dissipated, so many systems have a final planet at or around the 0.03 or 0.01 AU. This means that planets that form in the presence of a gas disc have semi-major axes that are not at all like that of the initial embryo distribution.

Mix.base and Frag.base produced an average of 2.80 planets per run. The Mix.wide and Frag.wide simulations were comparable, with an average of 2.84 planets per run. Both Mix.icy and Frag.icy show an inner population of final planets qualitatively similar those of Mix.base and Frag.base. Within 0.1 AU of the star, Mix.icy and Frag.icy had an average of 3.18 planets per run. However, at larger orbital distances, Mix.icy and Frag.icy average an additional 8.96 planets per run. Mix.edge and Frag.edge simulations experienced the most collisions, with an average of 1.51 planets per run. These values are also presented for each model in Table \ref{tab:planetsummary}.

A gas disc has a large impact on planet formation because it causes inward migration, resulting in more total collisions. Because the total mass was held constant in all of our models, more collisions correspond to fewer final planets. For all models, the minimum resulting mass of a body was simply the initial embryo mass, $0.08 \: \mathrm{M_{\oplus}}$. The embryos that avoid collisions for the entire simulation are typically initialized near the outer edge. Thus, the lack of collisions for these embryos can be attributed to edge effects and may not be physically meaningful. 

Mix.base and Frag.base produced planets with a median mass of $4.16 \: \mathrm{M_{\oplus}}$ and a maximum of $10.55 \: \mathrm{M_{\oplus}}$. Mix.wide and Frag.wide were again similar to the base models, with a median planet mass of $3.98 \: \mathrm{M_{\oplus}}$ and a maximum of $11.14 \: \mathrm{M_{\oplus}}$. Close-in super-Earths in this mass range have been observed; GJ 1214b is a $6.55 M_{\oplus}$ planet with a $1.58$ day orbital period orbiting an $0.16 M_{\odot}$ star \citep{Charbonneau_2009}. A recently-confirmed example is TOI-1235b, a $6.91 M_{\oplus}$ planet with a $3.44$ day orbital period orbiting an $0.64 M_{\odot}$ star  \citep{Cloutier_2020}.

We now consider the inner ($a<0.1$ AU) and outer ($a>0.1$ AU) planet populations for Mix.icy and Frag.icy separately. The inner population had a median mass of $4.14 \: \mathrm{M_{\oplus}}$ and a maximum of $9.63 \: \mathrm{M_{\oplus}}$, also comparable to the base models. The outer population, however, is composed primarily of small bodies, with a median mass of $0.08 \: \mathrm{M_{\oplus}}$ (i.e., an embryo that never experienced a collision) and a maximum of $0.72 \: \mathrm{M_{\oplus}}$. Mix.edge and Frag.edge produced planets with a median mass of $9.11 \: \mathrm{M_{\oplus}}$. In these models, the maximum planet mass was $11.76 \: \mathrm{M_{\oplus}}$, which is equivalent to the total amount of solids in the simulation (i.e., some simulations incorporated all the solids into just one planet). Models Mix.nogas and Frag.nogas produced planets with a median mass of $2.67 \: \mathrm{M_{\oplus}}$ and a maximum of $7.72 \: \mathrm{M_{\oplus}}$.

The median mutual inclinations for planets in Mix.base and Frag.base are $0.55^{\circ}$ and $1.01^{\circ}$, respectively. Because embryos experience damping in the presence of a gas disc, the median mutual inclination for planets in Mix.nogas ($7.44^{\circ}$) and Frag.nogas ($5.96^{\circ}$) are higher than that of most of the gaseous models. The exception is Mix.icy, which has a median mutual inclination of $8.65^{\circ}$. However, the inner ($a<0.1$ AU) planet populations for Mix.icy have a median mutual inclination of $0.68^{\circ}$, indicating that the high mutual inclinations in Mix.icy are largely driven by the outer planets.

The final dynamical separation is also largely dependent on the presence of gas during formation. Models including gas have a median dynamical separation of $\Delta = 13.47$, with $19.9\%$ closer than $10 R_{\mathrm{Hill}}$. Inward migration from gas torques result in dynamically closer planets, but most are dynamically isolated from each other. For gas-free models, the median dynamical separation between planets is $\Delta = 21.35$, with $0.15\%$ closer than $10 R_{\mathrm{Hill}}$.

Although nearly 1/5 of planets that formed in a gaseous environment have a final dynamical separation less than their isolation mass separation, we find that these systems are typically more stable than systems without gas. The stability of each system was evaluated using the AMD stability criterion \citep{Laskar_2017}. Systems that are AMD-stable may be considered long-term stable, while systems that are AMD-unstable require more robust analyses to determine stability. We find that at 100 Myr, 93.5\% of our base simulations are AMD-stable because the gas is effective at dissipating excess angular momentum after embryos collide. The percentage of AMD-stable systems in the wide and edge models are also high, 93.5\% and 99.0\%, respectively. In contrast, only 73.50\% of planetary systems that formed in our gas-free simulations are AMD-stable at 100 Myr.

In models including a gas disc, the disc begins rapid exponential dissipation at 4 Myr.  Systems Mix.base and Frag.base are AMD-stable 92.5\% of the time at 4 Myr, indicating that most systems stabilize in the presence of the gas disc. This trend carries over to the wide and edge models; at the onset of disc dissipation 90.5\% of the wide model systems are AMD-stable, and 99.0\% of the edge model systems are AMD-stable. Only 41\% of planetary systems that formed in a gas-free environment are AMD-stable at 4 Myr, though this value is only a point of comparison since there is no disc dissipation in these models.

The icy models are a notable exception to this trend. At disc dissipation, only 1 out of 50 Mix.icy systems is AMD-stable, and only 1 out of 50 Frag.icy systems is AMD-stable. At 100 Myr there is only a small increase in stability: 3 AMD-stable Mix.icy systems, and 5 Frag.icy systems. However, the lack of AMD stability is mostly driven by the population of less massive outer planets in these simulations. Considering only the inner ($a < 0.1$ AU) planets, 46 out of 50 Mix.icy systems are AMD-stable, and 45 out of 50 Frag.icy systems are AMD-stable at 100 Myr. If the system were to evolve beyond 100 Myr, it is possible that AMD from the outer system may be transferred to the inner system, potentially destabilizing the inner planets. However, the inner system might still avoid instability for billions of years before this occurs. Our solar system provides an example; if Jupiter's AMD were transferred to Mercury, Mercury and Venus would cross orbits despite having already survived for 4.5 Gyr. Future studies may probe how often this happens and on what timescales.

Since the presence of a gas disc causes planets to migrate, collisions are more likely to occur as orbits are pushed closer together. As a result, models with gas often have systems with fewer planets. The icy runs are an exception, as many of these planets did not spend enough time in the gas disc to migrate close to the star, and thus avoided some close encounters and collisions. Generally speaking, fewer remaining planets means there is less AMD in the system and less possible late-stage collisions. Models including a gas disc experience eccentricity and inclination dampening, which also remove AMD. The result is often a more stable system than if the gas had not been present.

It should be noted that disc migration often moves planets into mean motion resonances. Depending on the exact configuration of the planets, resonances can either stabilize or destabilize a system. The AMD stability criterion does not account for this effect (though it does account for resonance overlap, which is a separate source of instability), so AMD stability may characterize a system as stable even if planets are subject to resonant perturbations that destabilize the system. 

Therefore, the percentages of stable systems given above should be interpreted as an estimate of stability. In addition, collisions generally make a system more stable because the mean distance between planets increases after a collision and AMD typically decreases. Ejections stabilize the system by removing angular momentum, though in our simulations ejections were rare. After evolving for 100 Myr, most systems that experienced Type I migration tend to be more stable than systems that formed in the absence of gas. While the gas-free models experience no eccentricity or inclination damping, they may converge to the same level of stability as the gas models with additional integration time due to late collisions or ejections. Additionally, the icy models may experience some stabilizing late collisions if the system evolved past 100 Myr.

\subsection{Insensitivity to Initial Conditions}
\label{sec:results:inits}

\begin{figure*}
\centering
\includegraphics[width=\linewidth]{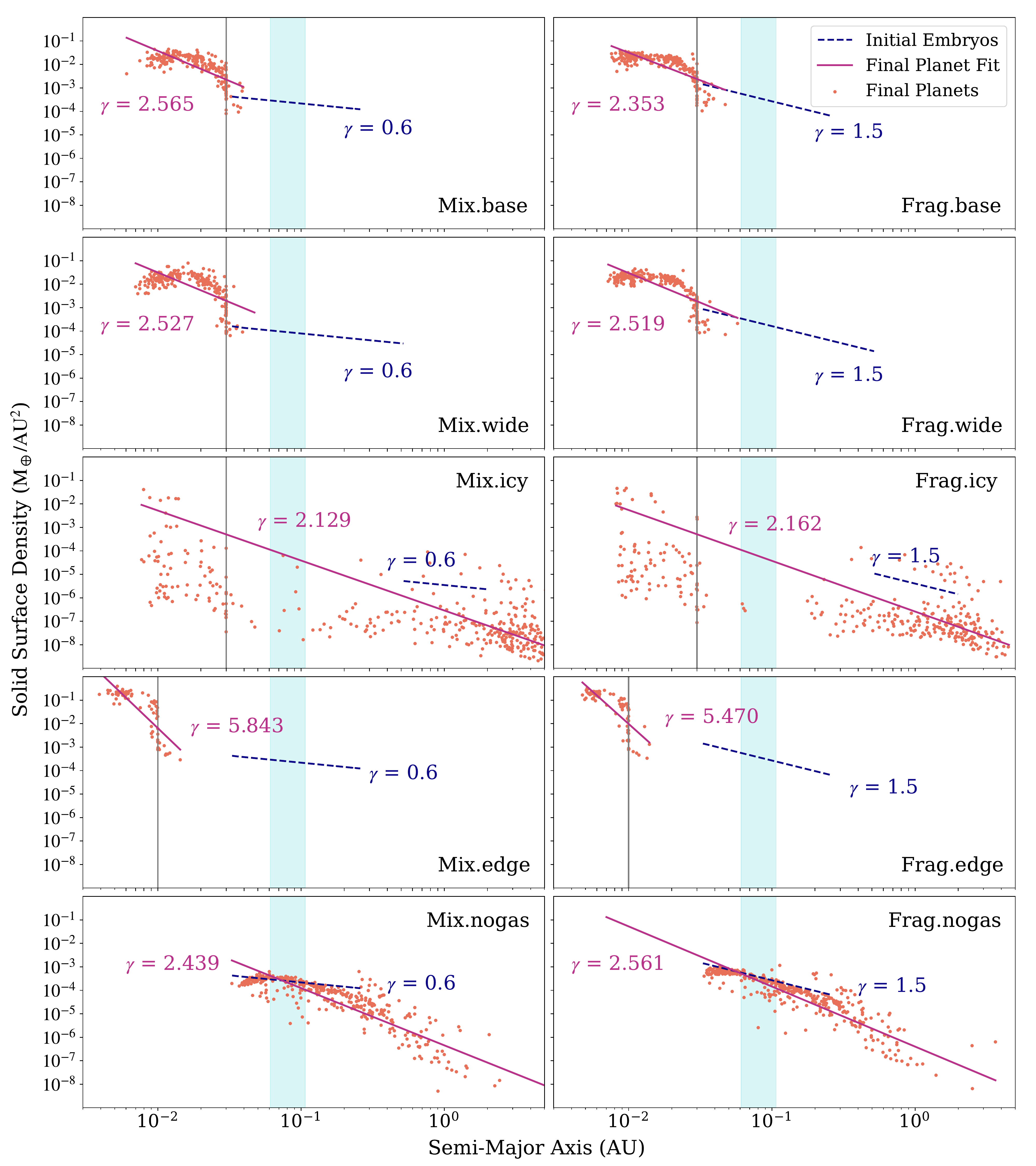}
\caption{Embryos are initialized according to a solid surface density profile that follows one of two uniform power laws (refer to Table \ref{tab:models}). We compare the initial density profile of the embryos (blue dashed line) to the final planets formed (orange points) and fit the final planets to a new power law (pink line) to guide the eye. The fitted power law index $\gamma$ for each model is also presented. We fit the final planets to a power law to provide a quantitative comparison of how the solid surface density changes, though a simple power law is not a good fit. Regardless of the absence or presence of a gas disc, the final density profile is completely different from the initial density profile of the embryos. For corresponding Mix and Frag models, $\gamma$ seems to tend toward a similar value regardless of the initial solid surface density profile. In other words, the final distribution of planets retains no memory of the initial solid surface density profile of the embryos.}
\label{fig:density}
\end{figure*}

Embryos were initialized according to the uniform power laws in Equations \ref{eqn:mixpl} and \ref{eqn:frag}. However, gravitational scattering, collisions, and migration redistribute the solids during the formation process. As a result, the solid surface density of the final system bears little resemblance to the initial distribution. Figure \ref{fig:density} compares the initial and final solid surface densities of each model. In order to compute a final surface density, we spread the mass of each planet into an annulus similar to the Minimum Mass Solar Nebula \citep{Hayashi_1981}. This allows us to obtain the average solid surface density profile around the planet's final semi-major axis.  We can then compare the initial density profile of the embryos to the final density profile of the resulting planets.

We find that for all models tested, the final solid surface density profiles no longer resemble the initial embryo distributions. This is true whether or not a gas disc is present; Figure \ref{fig:density} shows that all models produce planets with solid surface density profiles that are unrecognizable from those of the initial embryos. We computed the least squares fit of the final solid surface density profile to a power law. We find that corresponding Mix and Frag models tend to have similar final values of $\gamma$ despite having different initial values. Note that the solid surface density profile of the final planets are not a strong fit to a power law; we fit to a new uniform power law for the sole purpose of comparison.

We find that a broad range of initial conditions result in a population of planets at or near the location of the inner edge of the gas disc. However, individual embryos migrate depending on the strength of the gas torques. Because gas torques in the viscous regime are strongest close to the star (Figure \ref{fig:torques}), inner embryos tend to migrate inward faster than outer embryos. This is illustrated in Figure \ref{fig:semivstime}. Despite experiencing gravitational interactions and disc effects, embryos remain in approximately the same order. In other words, the outermost planets tend to form primarily from outer embryos, and the innermost planets tend to form primarily from inner embryos. 

While our results suggest that the formation process erases memory of the initial embryo solid surface density distribution, final planets do retain a memory of whether gas was present during formation. Models that include a gas disc typically produce a population of a few close-in planets because gas torques cause inward migration. In addition, planets retain some memory of where they formed relative to other planets in the system, as embryos tend not to significantly scramble their order during the formation process.

\section{Discussion}
\label{sec:discuss}

In Section \ref{sec:results} we presented our simulation results. Here, we discuss the probability of observing multiple planets in our model systems. We also compare the \textit{Kepler} sample of multi-planet systems to the systems that resulted from our simulations. Since \textit{Kepler} observed few true M stars, we construct a comparison sample that includes both M and K stars. Our sample contains all confirmed \textit{Kepler} multi-planet systems orbiting stars with mass $M_{\star} < 0.8 M_{\odot}$. We consider implications for atmosphere retention for planets similar to our models, briefly discussing several relevant processes. We also compare our simulated systems with a few observed M dwarf systems. Lastly, we detail the limitations of our models and identify parameters which could reasonably be changed.

\begin{figure}
\centering
\includegraphics[width=\linewidth]{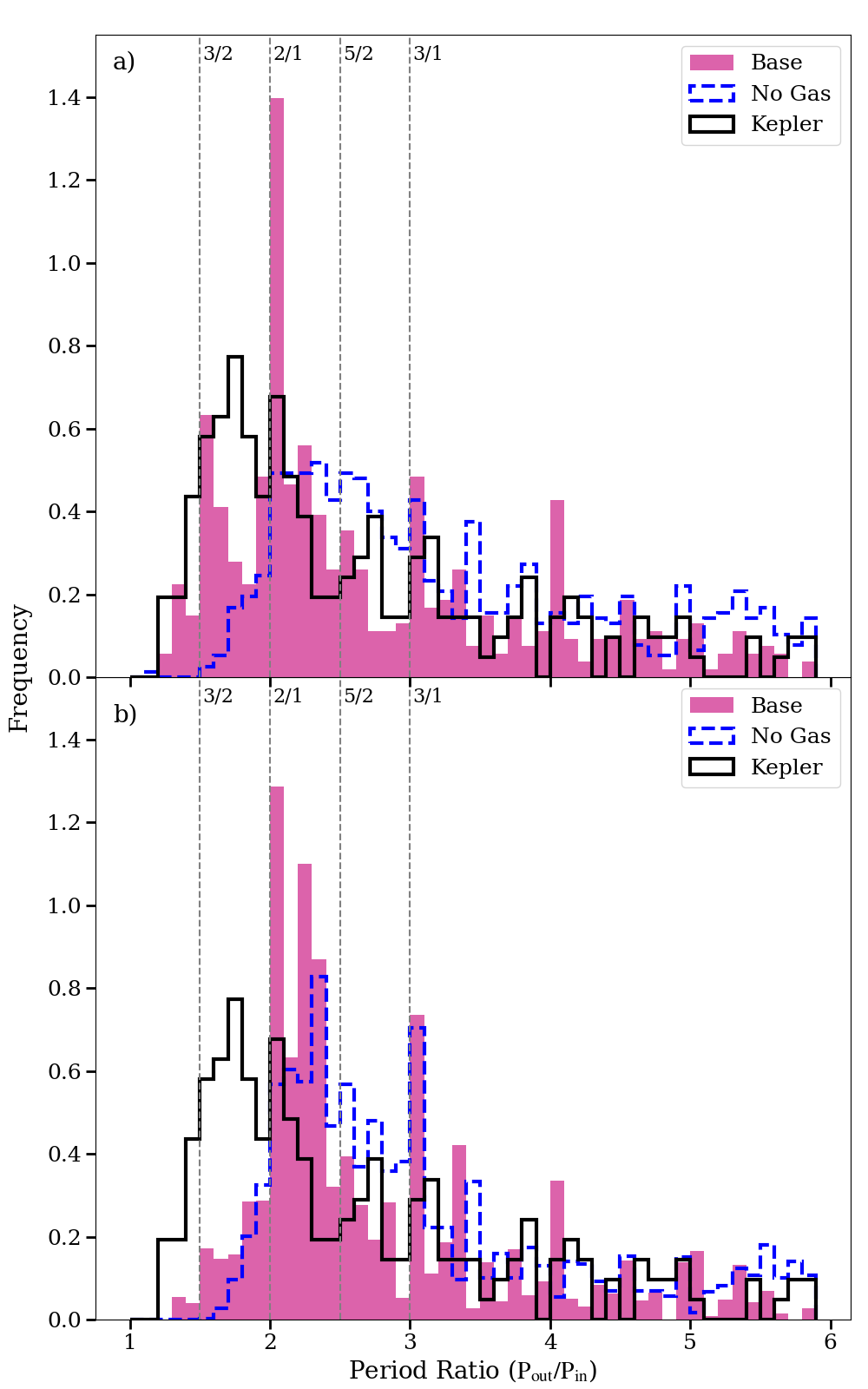}
\caption{A comparison of period ratios for Mix.base \& Frag.base (solid pink), Mix.nogas \& Frag.nogas (dashed blue line), and \textit{Kepler} data (black line). We restrict the \textit{Kepler} data to confirmed multi-planet systems orbiting stars with mass $M_{\star} < 0.8 M_{\odot}$. While multi-planet M dwarf systems have been found outside of \textit{Kepler}, here we use only \textit{Kepler} data to reduce biases. Panel a) displays period ratios for all simulated planets. Panel b) shows the period ratios weighted based on the geometric probability of both planets in a pair transiting. We use CORBITS to compute these probabilities \citep{Brakensiek_2016}.}
\label{fig:kepler}
\end{figure}

\subsection{Probability of Observation}
\label{sec:discuss:probs}

We present the typical mutual inclinations for gaseous and gas-free models in Section \ref{sec:results:planets}. \citet{Ragozzine_2010} give the probability of observing multi-transiting systems if one of the planets is known to transit, given the mutual inclination $\phi$ between planets and each planet's semi-major axis.
\begin{equation}
    p(1 \& 2 \mid 1) \simeq\left\{\begin{array}{ll}1 & \text { if } a_{1}>a_{2} \text { and } \phi \lesssim \frac{R_{*}}{a_{1}} \\ \frac{a_{1}}{a_{2}} & \text { if } a_{1}<a_{1} \text { and } \phi \lesssim \frac{R_{*}}{a_{1}} \\ \frac{R_{*}}{a_{2} \sin \phi} & \text { if } \phi \geq \frac{R_{*}}{a_{1}}\end{array}\right.
\end{equation}

We apply this formula to each of our final planetary systems to find the probabilities of observing a companion to a known transiting planet. Because most of our planets have relatively low final eccentricities (median over all runs: 0.03), we assume that the formula holds even though it is only strictly true for planets with circular orbits.

Given that one planet transits, the average probability of seeing an additional transiting planet is 64.4\% for our base models. Gas-free models (Mix.nogas \& Frag.nogas) tend to have larger semi-major axes and relatively high mutual inclinations, so the average probability of seeing an additional transiting planet decreases to 5.0\%. This has valuable implications for detecting multi-planet systems or planning RV follow-up on previously detected transiting planets. Because it is less likely that multiple planets transit in systems that formed in the absence of gas, our current transit observations of multi-planet systems may be biased toward systems that formed in a gaseous environment. Thus, planets that formed primarily without gas might require RV observations to detect other planets in the system.

Furthermore, these simulations suggest that a significant fraction of transiting planets could be accompanied by non-transiting planets that are detectable by RV follow-up campaigns. Placing an upper limit on the frequencies of such planets would constrain the frequency of formation in gas-free environments.

\subsection{Period Ratios}
\label{sec:discuss:comp}

We compute period ratios for our simulated population of planets,  comparing our base models with our gas-free models. We calculate the period ratio for each possible pair of planets in a system, not just adjacent planets. When comparing to observations, this choice is less sensitive to the mutual inclination distribution than using apparently adjacent planets. We do not modify the distribution to account for instrumental effects, but we weight our model period ratios according to their geometric transit probability for better comparison to the \textit{Kepler} sample. We compute these probabilities with CORBITS \citep{Brakensiek_2016}. The distribution of period ratios for the \textit{Kepler} sample, our base models, and our gas-free models are shown in Figure \ref{fig:kepler}.

After accounting for the probability that each simulated planet from our base models transits its host star, the peak that was previously at 3:2 disappears. This is because the simulated planets with period ratios around 3:2 tend to orbit at larger semi-major axes than planets with period ratios of 2:1 or higher. These planets, which formed in the presence of gas, often end up with a period ratio of 2, while most planets that formed in a gas-free environment have period ratios larger than this. The \textit{Kepler} sample has a broad distribution of period ratios qualitatively similar to both our base and gas-free models. The tail of the \textit{Kepler} distribution is qualitatively similar to those of our base and gas-free models. The period ratio distribution observed by \textit{Kepler} peaks at smaller period ratios than our models, even after we account for geometric transit probabilities. Our base models show a strong peak near 2:1, which is present but less pronounced in the \textit{Kepler} distribution \citet{Lissauer_2011}. After accounting for geometric transit probabilities with CORBITS, the \textit{Kepler} sample has a higher fraction of planet pairs around 3:2 compared to our models. However, we caution that the period ratio distribution observed by \textit{Kepler} is dominated by planetary systems around more massive stars. Our understanding of the period ratio distribution for M dwarf planets may improve as the number of multiple planet systems discovered around M dwarfs grows.

Our simulated sample of planets also shares the ``peas in a pod'' characteristics observed in the \textit{Kepler} sample \citep[e.g.][]{Weiss_2018, He_2019, He_2020}. That is, planets tend to be similar sizes as their neighbors, and adjacent planets tend to have period ratios that correlate with the period ratios of other adjacent planets in the system. Figure \ref{fig:planets} shows that this is qualitatively true for many simulations. Planets in multi-planet \textit{Kepler} systems tend to be about $20 R_{\mathrm{Hill}}$ apart and are rarely closer together than $10 R_{\mathrm{Hill}}$ \citep[e.g.][]{Lissauer_2011, Fabrycky_2014, Weiss_2018}. This is consistent with our simulated sample of planets, as presented in Section \ref{sec:results:planets}.

\subsection{Atmosphere Retention}
\label{sec:discuss:atmos}

Although the models presented here do not include gas accretion onto the planets or atmosphere formation, a key result of this work is that planets form long before the gas disc dissipates (Section \ref{sec:results:form}). A natural follow-up question to this result is whether these planets could have significant atmospheres, given that they form almost entirely in a gaseous environment. In this section we briefly discuss atmosphere retention to address this point and speculate on how various mechanisms could remove or stop the growth of atmospheres around our simulated planets.

\citet{Owen_2013} find that photoevaporation from a Sun-like star can remove massive hydrogen atmospheres inward of approximately 0.1 AU for Neptune-mass objects. They find that planets must be made of rocky cores of a few Earth masses in order to reproduce the observed features in the \textit{Kepler} sample. These values do not generalize for M dwarfs, but provide a starting point for considering the potential retention of atmospheres for close-in M dwarf planets.

In our models, planets often migrate to the inner edge of the gas disc or even further. If a planet accretes a gaseous envelope before migrating into the disc cavity, then its envelope could be photoevaporated to leave behind a rocky core. \citet{Lopez_2013} emphasize the importance of core mass when considering effects of photoevaporation. Planets that formed in our base models had a median mass of approximately $4 M_{\oplus}$. These low core masses further suggest that photoevaporation would reduce the likelihood of a planet retaining its atmosphere. Envelopes accreted in a gas disc may have cooling times that are significantly longer than the migration timescale, meaning the atmospheres have not yet cooled and contracted before being subjected to the effects of photoevaportation \citep{Owen_2016}. This makes it easier for the planetary atmosphere to be removed, which may help explain the radius gap in the \textit{Kepler} sample \citep{Fulton_2017, Owen_2017}. While the radius gap is well-established for more massive stars, for M stars there is no observational evidence for a radius gap \citep{Hsu_2020}.

Another process that could remove planet atmospheres is core-powered mass loss, during which the heat released during formation evaporates the atmosphere \citep{Ginzburg_2018}. Unlike photoevaporation of atmospheres, core-powered atmospheric mass loss does not depend on stellar irradiation. Instead, the amount of mass lost depends on the equilibrium temperature of the planet. Simulations of core-powered mass loss from \citet{Loyd_2020} produce a radius gap that is comparable to the the radius gap observed in the data, differing only in location. This may be explained by differences in stellar mass, though they find no stellar mass dependence on the exoplanet radius gap of their sample. While both photoevaporation and core-powered mass loss are proposed mechanisms to explain the radius gap around Sun-like stars, it is currently uncertain whether one process dominates over the other and whether the dominant mechanisms could depend on host star properties. Further research is needed to determine the relative importance of these processes around M dwarfs.

Additionally, \citet{Kite_2019} propose that planet growth could be stopped by mantle-atmosphere interactions. At high atmospheric pressures, hydrogen becomes soluble in magma, allowing atmospheric hydrogen to be incorporated into the core. Models show that this effect can reproduce the cliff around $3 R_{\oplus}$ in the planet radius distribution. However, further research is needed to better understand atmospheric mass loss mechanisms and their relative importances.

\subsection{Comparison to Known Systems}
\label{sec:discuss:trappist}

In this study we do not attempt to reproduce any specific known planetary systems. Rather, we consider the generic outcome of planet formation for relatively close-in solids (no further than 0.52 AU at a disc age of 1 Myr) around a 0.2 $\mathrm{M_{\odot}}$ star. Here, we compare the outcomes of our models to a few noteworthy M dwarf systems, though the list is not exhaustive.

The TRAPPIST-1 system is a popular and well-studied multiplanet M dwarf system \citep{Gillon_2016,Gillon_2017}. Recent estimates for TRAPPIST-1 find that its mass is $M_{*}=0.089 \pm 0.006 M_{\odot}$, less than half the mass of the host star in our models \citep{VanGrootel_2018}. The TRAPPIST-1 system is notable for its 7 close-in planets, which have semi-major axes ranging from roughly 0.012 - 0.062 AU \citep{Grimm_2018}. While our gaseous models produce fewer close-in planets per system, planets do migrate into similar semi-major axes as the TRAPPIST-1 planets. Close comparison with our models may not be meaningful as this work was not done with any particular system in mind, however, other studies attempt to explain the formation of TRAPPIST-1
\citep[e.g.][]{Ormel_2017, Schoonenberg_2019, Coleman_2019}.

Teegarden's Star ($M_{*}=0.089 \pm 0.009 M_{\odot}$) hosts two known planets \citep{Schweizer_2019,Zechmeister_2019}. The planets have a period ratio of roughly 2.3 and orbit at 0.0252 AU and 0.0443 AU for planets b and c, respectively. These orbital characteristics closely resemble systems resulting from several of our gaseous runs. In our simulations, the outermost planet often ended up at the inner edge of the gas disc. It is possible that the same process occurred during the formation of the Teegarden's Star system.

Proxima Centauri, our nearest stellar neighbor, is an M dwarf of mass $M_{*}=0.122 \pm 0.002 M_{\odot}$ \citep{Mann_2015}. It hosts two known planets, with the Proxima b at 0.0485 AU and Proxima c at 1.48 AU \citep{AngladaEscude_2016,Damasso_2020}. Because our work focuses on close-in planets, none of our simulations result in planets that orbit at distances comparable to Proxima c.

\subsection{Model Limitations}
\label{sec:discuss:lims}

\subsubsection{Initial Conditions}
Future studies could consider exploring alternative choices for many of the initial conditions in this study. For example, in the models presented here we primarily initialize embryos in the inner disc ($0.033 < a/\rm{AU} < 0.26$), though some of our models extend out further (0.52 AU and 2.0 AU). The initial embryo distribution in real systems is unknown, and this choice allowed us to focus on the behavior of young planetary systems in a disc dominated by viscous heating.

Additionally, we assume the inner edge of the disc to be 0.03 AU in most runs. Models Mix.edge and Frag.edge assume a disc inner edge of 0.01 AU. However, these are only approximations for a 0.2 $\mathrm{M_{\odot}}$ star. The precise inner edge of the disc is not well-constrained, as it depends on stellar activity and and stellar rotations of pre-main sequence low-mass stars, which are also poorly constrained. We chose the value for the $\alpha$-viscosity of the disc to be $\alpha=10^{-3}$, which is generally consistent with the literature \citep{Birnstiel_2012, Ida_2016}. Because the characteristics of a typical disc around a 0.2 $\mathrm{M_{\odot}}$ star are somewhat uncertain, however, different values of $\alpha$ could reasonably be tested.

The age at which planetesimals form could also be adjusted. In this study we assume that planetesimals form by $t = 0.1$ Myr, though recent work suggests that planetesimals may form even sooner \citep[e.g.][]{Lenz_2019}. Planetesimals forming earlier in the disc lifetime would increase the total amount of mass available for planet formation, which would result in more massive final planetary systems and further reinforce our conclusion that planet formation around M dwarfs occurs very rapidly.

In all models, we initialize embryos of equal mass according to the specified solid surface density power law. This means that the embryos were not spaced equally and were dynamically closer together at larger initial semi-major axes (Figure \ref{fig:dynamical}). However, we also tested an alternative case where embryos are equally spaced in Hill radii and the embryo masses were allowed to vary. Although we do not provide an in-depth discussion of these equal-distance embryo models, we note that they yield the same major conclusions as the equal mass embryo models presented in this work.

\subsubsection{Lifetime of Disc}
The results are fairly insensitive to the lifetime of the gas disc, as planets mostly form within the first 1 Myr. Since a typical disc lifetime is on the order of a few Myr, we expect that any reasonable choice of disc lifetime would yield similar results. In this work, we assume planetesimals form early in the disc lifetime, allowing them to experience disc effects like gas torques for 4 Myr. If planetesimals formed significantly later, that would prevent planets from migrating so far inward, as each planetesimal would feel gas torques for a shorter amount of time. Thus, varying the time when planetesimals form relative to the age of the disc may significantly alter the characteristics of the resulting planetary system, including location and stability. However, such drastic changes to these timescales are not physically motivated.

Assuming planetesimals form earlier would also increase the amount of available gas mass in the disc. We use the Toomre Q parameter to consider gravitational instability of the disc as the amount of available mass increases \citep{Toomre_1964}. A Keplerian disc is unstable when
\begin{equation}
    Q \equiv \frac{\Omega c_{s}}{\pi G \Sigma} < 1.
\end{equation}

\noindent For our disc at 1 Myr, $Q > 1$ for approximately $r < 1000 \; \rm{AU}$. If we increase $\Sigma$ by a factor of 100, which corresponds roughly to a disc age of $10^{4}$ yr, then $Q < 1$ closer than 1 AU. At this disc age the mass accretion rate onto the star is higher, so we expect this region of instability to be relatively short-lived, stabilizing well before planetesimals form.

\subsubsection{Planet Traps}
\label{sec:discuss:traps}

Planet migration traps occur when a region of the disc has a steep temperature gradient ($d \ln T / d \ln r < -0.87$) and the co-rotation torque overwhelms the inward Lindblad torque \citep{Brasser_2017}. In our disc model, that never occurs and migration is always inward. A similar work by \citet{Miguel_2020}, with a very similar disc model and stellar mass, does have planet traps. The difference occurs because they take into account changes of dust opacity at the snow line, and also because they assume a fixed $\dot{M}$ that results in a cooler disc, which brings the snow line into the viscous region of the disc. In our model we take into account the evolution of the accretion rate and stellar luminosity. The result is that the snow line stays out of the viscous region.

If we had treated the snow line as in \citet{Miguel_2020}, our model still would not have formed a planet migration trap. If M dwarf discs typically have planet migration traps, our main result (e.g., that planet formation is rapid) would not change, but the final planet orbits would be further out than in our model. We do not see the final orbits of our simulation as a model prediction, as they mainly reflect the location of the inner edge of the disc which is largely unknown for discs around M dwarfs.

\subsubsection{Type II Migration}
\label{sec:discuss:t2}

In this work we consider only the effects of Type I migration. However, some planets in our simulations likely grow large enough to open a gap in the disc and experience Type II migration. Using the gas mass fraction formula from \citet{Ginzburg_2016}, we find that the most massive planets in our simulations ($10.84 M_{\oplus}$) may accrete approximately $36\%$ of their mass in gas before disc dispersal. More typical simulated planets (around $4 M_{\oplus}$) may accrete around $16\%$ of their mass in gas by disc dispersal. In either case, such small gas fractions will have a negligible effect on the orbital dynamics and only a small impact on the planet's Hill radius.

Computing the disc scale height from \citet{Ida_2016} and applying the gap opening criterion from \citet{Crida_2006}, we find that planets around 0.1 AU from the star may start to open gaps after growing to $4 M_{\oplus}$ or larger. When planets open gaps, they undergo Type II migration and the migration speed greatly decreases. As a result, our simulations may overestimate the migration rate for the more massive planets near the end of the run. We do not expect this to significantly affect any of our key results since most planets only become massive enough to potentially open a gap once they have already migrated close to their final orbits. The rest would still be expected to migrate to the edge of the disc on the viscous timescale. The key results of our investigation --- that planet assembly around M dwarfs is extremely rapid, and that the final planetary system retains little memory of its initial conditions --- should not be sensitive to a last minute switch from Type I to Type II migration. Including Type II migration would not have a large impact on the final orbits of the planets in our simulations.

\subsection{Comparison to Previous Studies}

Considerable efforts have already been made in probing the formation of super-Earths. Much like this study, \citet{Ogihara_2015} examine the formation of close-in super-Earths via N-body simulations with and without gas disc effects (i.e., Type I migration and eccentricity/inclination damping), finding that Type I migration can cause planets to form in only a few Myr. They find that around G dwarfs, the presence of disc effects determine the final orbital configurations of the planets. As there is currently no known mechanism that suppresses Type I migration, a gas disc must be included in simulations of close-in planet formation to ensure that the model is physically meaningful. The models presented in this study examine the effects of a gas disc on the formation of similar planets (close-in super-Earths), except focused around M dwarfs.

Some studies have recently begun exploring the formation process of M dwarf planets. Many of these studies (including the work presented here) use N-body simulations plus approximations for additional effects such as migration in order to model the formation process. Others use population synthesis models that focus on pebble or planetesimal accretion. Here we briefly survey recent studies focusing on planet formation around low-mass stars.

\citet{Alibert_2017} use N-body formation models and gas accretion calculations with equal-mass embryos to predict the radius distribution and water content of planets around 0.1 $M_{\odot}$ stars. They consider 10 Moon mass embryos drawn from a log-uniform distribution between 0.01 and 5 AU, while our work arranges 147 embryos, each $0.08 M_\oplus$ ($\sim 6.5$ Moon mass), according to one of four specified solid surface density power laws. Like this study, the simulations begin at a protoplanetary disc age of 1 Myr.

\citet{Coleman_2019} compare planet formation via N-body pebble accretion and planetesimal accretion for the TRAPPIST-1 system. Simulations begin with 30 $0.08 M_\oplus$ embryos placed between 1 and 5 AU around a 0.1 $M_{\odot}$ star. For the planetesimal accretion scenario, 2000 planetesimals of varying masses were also placed between 0.5 and 5.5 AU. They find that TRAPPIST-like systems readily form in either accretion scenario and observational differences between scenarios are minor. In contrast, this study aims to explore the predictions of planet formation models around low-mass stars more generally rather than attempting to produce a TRAPPIST-like system.

\citet{Miguel_2020} use a population synthesis code to model planet formation around low-mass stars. They consider planetesimal accretion and include calculations for both type I and II migration. Like this study, they follow \citet{Ida_2016} to inform disc parameters. Their simulations are unable to form planets comparable to some of the most massive observed M dwarf planets, such as LHS1140b, GJ1214b, GJ3323c, and GJ 3512b. This suggests that M dwarf planet formation may be more efficient than previously expected, supporting the efficient formation process seen in our simulations.

\citet{Liu_2019} examine the formation of super-Earths using a pebble-driven core accretion model around a variety of stellar masses, including masses in the M dwarf range. They use Monte Carlo sampling of the initial conditions to model the evolution of the systems, and differentiate between formation models where embryos either all originate at the water ice line or are log-uniformly distributed across the protoplanetary disc. All their simulations result in a single-planet system. Using similar methods, \citet{Liu_2020} investigate the formation of single-planet systems around low-mass stars. They find that pebble accretion may allow protoplanets around a 0.1 $M_{\odot}$ star to grow to up to 2-3 $M_{\oplus}$, though most do not accrete this much mass. In contrast, the models presented in this study do not restrict the final number of planets in a system, and find that nearly all of the simulations (>99\%) with the inner edge of the gas disc at 0.03 AU result in multi-planet systems. For simulations with the inner edge of the gas disc at 0.01 AU, the percentage of runs with multi-planet systems decreases to 51\%.

While this study examines the formation of super-Earths through the accretion of embryos, it is worthwhile to acknowledge that there may also be other formation mechanisms at play. The recent discovery of GJ 3512b, a 0.46 $M_{\rm{Jup}}$ planet orbiting a 0.123 $M_{\odot}$ star, suggests that disc instability may be a favorable mechanism for giant planet formation around M dwarfs \citep{Morales_2019}.

\subsection{Future Work}

Future investigations should expand on the range of M dwarf masses modeled, as M dwarfs span nearly a full order of magnitude in mass. In addition, it may be helpful to consider alternative ranges of initial embryo semi-major axes, especially those that extend beyond the ice line. Simulations that encompass the ice line could also have implications for volatile delivery, such as the work presented in \citet{Ciesla_2015}. It may also be useful to explore whether final solid surface density could be used as a probe of the initial conditions. Furthermore, future simulations should evolve systems for longer than 100 Myr to explore whether late collisions continue to occur. Future work should consider the distribution and mixing of rocks and ices during the formation process in order to investigate the final composition of planets. An expanded model would require the consideration of both viscous and irradiative heating regimes, whereas in this work we initialized embryos according only to the power law implied by viscous heating. Lastly, it would be useful to further explore the potential role of planet traps, testing a variety of disc models and examining under what conditions and where traps arise. The presence of planet traps likely alters the final distribution of planets, so it would be useful to better understand the disc conditions that result in such traps.

\section{Conclusions}
\label{sec:conclude}

Our simulations lead us to four main conclusions:
\begin{itemize}
    \item Planets form quickly around $0.2 M_{\odot}$ stars, long before the gas disc dissipates. When planets become locked into resonant chains, they can be pushed far into the disc cavity via Type I migration of the outermost planet in the chain.
    
    \item Planet formation reshapes the solid distribution and destroys memory of initial conditions. The solid surface density of the final planets does not appear to be related to the initial distribution of embryos. Thus, it may not be possible to infer the initial distribution of solids from present-day observations of planetary systems.
    
    \item The location and number of planets in a system may reflect whether gas was present during formation. Systems that formed with gas tend to have a few close-in planets that are closely spaced, while systems that formed without gas tend to have more planets that are further out and more dynamically separated. In addition, gas-free systems tend to have smaller mutual inclinations than gas systems.
    
    \item While many late-stage giant impacts occur in the disc cavity or after disc dissipation, approximately a quarter of runs including a gas disc had a final giant impact inside the disc. Because these planets largely form inside the gas disc, they may retain an atmosphere even after migrating into the disc cavity.
    
\end{itemize}

\noindent We also find that systems that form in the presence of gas tend to be more stable than those that did not, according the the AMD stability criterion from \citet{Laskar_2017}. It is unclear whether this would remain true if we ran simulations out to 1 Gyr or longer, but gas systems certainly become AMD-stable much faster than gas-free systems. Though the icy models formed planets in the presence of gas, these systems are typically not AMD-stable due to the population of small bodies at larger orbital distances.

The distribution of \textit{Kepler} M dwarf period ratios peaks at slightly smaller period ratios than our base and gas-free period ratio distributions. Future models that span a larger range of stellar masses combined with new \textit{TESS} observations will further develop our understanding of planet formation around M dwarfs.

\section*{Acknowledgements}
\label{sec:ack}
The authors acknowledge suggestions for improving the clarity and organization of the manuscript by an anonymous referee.   
B.Z. and E.F. were supported in part by NASA grant NNX16AB44G. D.C.’s research was supported by an appointment to the NASA Postdoctoral Program within NASA’s Nexus for Exoplanet System Science (NExSS), administered by Universities Space Research Association under contract with NASA. D.C. acknowledges support from NASA Exoplanet Research Program award NNX15AE21G. 
This work was supported by a grant from the Simons Foundation/SFARI (675601, E.B.F.).  
E.B.F. acknowledges the support of the Ambrose Monell Foundation and the Institute for Advanced Study.
The Center for Exoplanets and Habitable Worlds is supported by the Pennsylvania State University, the Eberly
College of Science, and the Pennsylvania Space Grant Consortium. This research or portions of this research were conducted with Advanced CyberInfrastructure computational resources provided by The Institute for Computational \& Data Science at The Pennsylvania State University (http://ics.psu.edu), including the CyberLAMP cluster supported by NSF grant MRI-1626251. The Center for Exoplanets and Habitable Worlds is supported by the Pennsylvania State University, the Eberly College of Science, and the Pennsylvania Space Grant Consortium. The authors would like the thank Edwin Kite for his comments on volatiles and habitability, and Bertram Bitsch for the helpful discussion of planet migration.

\section*{Data Availability}
\label{sec:dataavail}
The data underlying this article will be shared on reasonable request to the corresponding author.


\bibliographystyle{mnras}
\bibliography{main}



\appendix



\section{Disc torques}
\label{app:torques}

This appendix is a slightly modified version of Appendix A of \citet{Carrera_2019} with the correction of a typo in Equation \ref{eqn:damp_forces}. Following \citet{Paardekooper_2010,Paardekooper_2011}, all the formulas below include a smoothing of the planet potential with a smoothing length of $b = 0.4h$ where $h = H/r \approx 0.05$. In Type-I migration, a planet experiences a negative Lindblad torque $\Gamma_{\rm L}$ and a positive co-rotation torque $\Gamma_{\rm C}$. The total torque on the planet is given by

\begin{equation}\label{eqn:Gamma_tot}
	\Gamma_{\rm tot} = \Gamma_{\rm L} \Delta_{\rm L}
    				 + \Gamma_{\rm C} \Delta_{\rm C}
\end{equation}

The expressions for $\Delta_{\rm L}$ and $\Delta_{\rm C}$ were produced by \citet{Cresswell_2008,Coleman_2014,Fendyke_2014} and collected by \citet{Izidoro_2017},

\begin{eqnarray}\nonumber
    \Delta_{\rm L}
 	&=&
    \left[
    P_{\rm e} + \frac{P_{\rm e}}{|P_{\rm e}|} \times
    \left\lbrace
    	0.07 \left( \frac{i}{h}\right)
    	+ 0.085\left( \frac{i}{h}\right)^4\right.\right.\\
    && \left.\left.
        - 0.08\left(  \frac{e}{h} \right) \left( \frac{i}{h} \right)^2
	\right\rbrace
    \right]^{-1}\\
	\Delta_{\rm C}
    &=&	\exp\left( \frac{e}{e_{\rm f}} \right)
    	\left\lbrace
        	1 - \tanh\left(\frac{i}{h} \right)
        \right\rbrace
\end{eqnarray}

where $e$ and $i$ are the planet orbital eccentricity and inclination, where $e_{\rm f} = 0.5h + 0.01$, and 

\begin{equation}
	P_{\rm e}
    = \frac{
    	1 + \left( \frac{e}{2.25h}\right)^{1.2} + \left( \frac{e}{2.84h}\right)^6
	}{
    	1 - \left( \frac{e}{2.02h}\right)^4
	}.
\end{equation}

The expressions for $\Gamma_{\rm L}$ and $\Gamma_{\rm C}$ were derived by \citet{Paardekooper_2010,Paardekooper_2011}

\begin{eqnarray}
	\frac{\Gamma_{\rm L}}{\Gamma_0/\gamma_{\rm eff}}
    &=& (-2.5 -1.7\beta + 0.1x),\\
    \nonumber
	\frac{\Gamma_{\rm C}}{\Gamma_0/\gamma_{\rm eff}}
    &=&   1.1\left( \frac{3}{2}-x\right) F(p_\nu)~G(p_\nu)\\
    \nonumber
	&+&   0.7\left( \frac{3}{2}-x\right) (1 - K(p_\nu)) \\
    &+&     \frac{7.9 \xi}{\gamma_{\rm eff}} F(p_\nu) F(p_\chi)
            \sqrt{G(p_{\rm \nu}) G(p_\chi)}\\
	&+&     \left( 2.2 - \frac{1.4}{\gamma_{\rm eff}}\right)\xi
    		\sqrt{(1 - K(p_\nu))~(1 - K(p_\chi)}\nonumber
\end{eqnarray}

where $\xi = \beta - (\gamma -1)x$ is the negative of the entropy slope, $\gamma = 1.4$ is the adiabatic index, $x$ is the negative of the surface density profile, $\beta$ is the temperature gradient, and $\Gamma_0$ is a scaling factor,

\begin{equation}
	x = - \frac{\partial {\rm ln} ~\Sigma_{\rm gas}}{\partial {\rm ln}~r},
	~~~
    \beta = - \frac{\partial {\rm ln} ~T}{\partial {\rm ln}~r}.
	~~~
	\Gamma_0 = \left(\frac{q}{h}\right)^2 \Sigma_{\rm gas} r^4 \Omega_k^2,
\end{equation}
and $q$ is the planet-star mass ratio. Next, we define the effective adiabatic index $\gamma_{\rm eff}$

\begin{eqnarray}
	\gamma_{\rm eff}
    &=&
    \frac{2~Q \gamma}{
    	\gamma Q + \frac{1}{2}
        \sqrt{2 \sqrt{( \gamma^2Q^2+1)^2-16 Q^2(\gamma-1)}+2 \gamma^2 Q^2 - 2}
    },\\
	Q
    &=&
    \frac{2 \chi}{3 h^3 r^2 \Omega_k},\\
	\chi
    &=&
    \frac{16~\gamma~(\gamma -1)~\sigma T^4}{3 \kappa~\rho^2~(hr)^2~\Omega_k^2},
\end{eqnarray}
where $\rho$ is the gas volume density, $\kappa$ is the opacity and $\sigma$ is the Stefan-Boltzmann constant. Next, $p_\nu$ and $p_\chi$ are parameters that govern the viscous saturation and the thermal saturation respectively. They are defined in terms of the non-dimensional half-width of the horseshoe region ${x_{\rm s}}$,

\begin{equation}
	p_\nu = \frac{2}{3}\sqrt{\frac{r^2\Omega_k}{2\pi\nu}x_s^3},
	~~~
	p_\chi = \sqrt{\frac{r^2\Omega_k}{2\pi\chi}x_s^3},
	~~~
	x_{\rm s} = \frac{1.1}{\gamma_{\rm eff}^{1/4}}\sqrt{\frac{q}{h}}.
\end{equation}

Finally, we give the expression for functions $F$, $G$, and $K$,

\begin{eqnarray}
  F(p)
  &=& \frac{1}{1+ \left( \frac{p}{1.3}\right)^2}\\
  G(p)
  &=&
    \left\{
        \begin{array}{ll}
            \frac{16}{25}
            \left( \frac{45\pi}{8}\right)^{3/4} p^{3/2}
            & {\rm if}~~ p < \sqrt{\frac{8}{45\pi}}\\
            1 - \frac{9}{25}
            \left( \frac{8}{45\pi}\right)^{4/3} p^{-8/3}
            & {\rm otherwise}
        \end{array}
    \right.\\
  K(p)
  &=&
    \left\{
        \begin{array}{ll}
            \frac{16}{25}
            \left( \frac{45\pi}{28} \right)^{3/4} p^{3/2}
            & {\rm if}~~ p < \sqrt{\frac{28}{45\pi}}\\
            1 - \frac{9}{25}
            \left( \frac{28}{45\pi} \right)^{4/3} p^{-8/3}
            & {\rm otherwise}\label{eqn:Kp}
        \end{array}
    \right.
\end{eqnarray}

Equations \ref{eqn:Gamma_tot}-\ref{eqn:Kp} are enough to compute the total torque $\Gamma_{\rm tot}$. To implement disc torques as an external force in an N-body code, we follow \citet{Papaloizou_2000} and \citet{Cresswell_2008} in defining the planet migration timescale as $t_{\rm mig} =- L / \Gamma_{\rm tot}$, where $L$ is the planet's orbital angular momentum. The force term becomes,

\begin{equation}\label{eqn:a_mig}
	\mathbf{a}_{\rm mig} = -\frac{\mathbf{v}}{t_{\rm mig}},
\end{equation}
where $\mathbf{v}$ is the planet's instantaneous velocity. To implement eccentricity and inclination damping, we need their damping timescales ${\rm t_e}$ and ${\rm t_i}$. Expressions for these timescales have been derived by \citet{Papaloizou_2000} and \citet{Tanaka_2004}, and were later modified by \citet{Cresswell_2006,Cresswell_2008},

\begin{eqnarray}
	t_e
    &=&
    \frac{t_{\rm wave}}{0.780}
    \left[
    	1-0.14\left(\frac{e}{h}\right)^2 
        + 0.06\left(\frac{e}{h}\right)^3\right.\\
    &+&\left.
        0.18\left(\frac{e}{h}\right)\left(\frac{i}{h}\right)^2
    \right]\nonumber\\
	t_i
    &=&
    \frac{t_{\rm wave}}{0.544}
    \left[
    	1-0.3\left(\frac{i}{h}\right)^2 
        + 0.24\left(\frac{i}{h}\right)^3\right.\\
    &+&\left.
        0.14\left(\frac{e}{h}\right)^2\left(\frac{i}{h}\right)
    \right]\nonumber\\
	t_{\rm wave}
    &=&
    \left(\frac{M_{\odot}}{m}\right)
    \left(\frac{M_{\odot}}{\Sigma_{\rm gas} a^2}\right)
    h^4 \Omega_k^{-1},
\end{eqnarray}
where $m$ and $a$ are the planet's mass and semimajor axis. Similar to Equation \ref{eqn:a_mig}, we implement eccentricity damping and inclination damping, respectively, as the forces

\begin{equation}\label{eqn:damp_forces}
	\mathbf{a}_e = -2~\frac{(\mathbf{v \cdot r})\mathbf{r}}{r^2 t_e},
	~~~
	\mathbf{a}_i = -\frac{2 v_z}{t_i}\mathbf{k}
\end{equation}

where $\mathbf{r}$ and $\mathbf{v}$ are the position and velocity vectors of the planet, $v_z$ is the $z$ component of the planet's velocity, and $\mathbf{k}$ is the unit vector in the $z$ direction.


\bsp	
\label{lastpage}
\end{document}